\documentclass[twocolumn]{aa}
\usepackage{graphicx}
\usepackage{txfonts}
\begin{document}
   \title{The Luminosity Function of Cluster Galaxies. III.}

%   \subtitle{}

   \author{I. Parolin
          \inst{1}
          \and
      E. Molinari
          \inst{1}
      \and
      G. Chincarini
      \inst{1}\fnmsep\inst{2}
          }

   \offprints{I. Parolin,
   \email{parolin@merate.mi.astro.it}}

   \institute{INAF - Osservatorio Astronomico di Brera, Via Bianchi 46, 23807 Merate (LC), Italy
         \and
             Universit\`a degli Studi di Milano-Bicocca, Piazza dell'Ateneo Nuovo 1, 20126, Milano, Italy
                     }

   \date{Received March 12, 2003; accepted June 09, 2003}

   \abstract{
We investigated the optical properties of 7 clusters of galaxies observed in three colors over the range of absolute magnitudes $ -24 \leq {\rm M} \leq -12$.  Our aim is to estimate the Luminosity Function and the total cluster luminosity of our sample in order to have information about the formation and the evolution of galaxies in clusters. In this paper, we present the main points of our analysis and give the formal parameters obtained by fitting the data using the maximum likelihood algorithm. We find consistency between our results and other works in literature confirming the bimodal nature of the luminosity function of cluster galaxies. More important, we find that the relation $ {\rm L_{opt}/ L_{X}}$ versus ${\rm L_{X}}$ is color dependent: Low X ray Luminosity clusters have a bluer galaxy population.

   }

   \maketitle
%
%----------------------------------------------------------------------------
\section{Introduction}

The estimate of the total cluster luminosity in different colors, the estimate of its Luminosity Function (LF) and the comparison with the LF of field galaxies (see also \cite{Chi1988}) are powerful tools to gather information on how a cluster of galaxies forms, evolves and accretes from the field.

Currently, we have at least two determinations of the form of the local field LF (Binggeli et al. 1988) which are in excellent agreement: \cite{Zu1997} and \cite{Bl2001}. \cite{Bl2001} estimate the local field LF in different colors down to an absolute magnitude ${\rm M \sim -16}$ (Petrosian magnitudes) using the Sloan Digital Sky Survey (SDSS) commissioning data. The very large sample they used allowed them to have very good statistics and they estimated:  ${\rm \phi = (1.46 \pm 0.12)\cdot 10^{-2} h^{3} Mpc^{-3}}$,  ${\rm M^{*}(r) = -20.83 \pm 0.03}$, ${\rm \alpha = -1.20 \pm 0.03}$.  These values are comparable with the ones found by \cite{Zu1997}, who analysed the local field LF down to magnitudes ${\rm M \sim -12}$ through the ESO Slice Project (ESP) survey: ${\rm \phi = (0.020 \pm 0.004)h^{3} Mpc^{-3}}$,  ${\rm M^{*}(Bj) = -19.61 \pm 0.07}$, ${\rm \alpha = -1.22 \pm 0.06}$.  The data from the ESP survey, four magnitudes deeper than SDSS, showed a steepening of the faint-end slope starting at magnitude ${\rm M_{Bj} \sim -17}$ that Zucca et al. fitted with a power law with slope ${\rm \beta = -1.6}$. In spite of the small number of objects with ${\rm M_{Bj} \ge -16 }$, this is a very important result, as it is in line with the value found by \cite{Dri1994}, and it is in contrast with the value that \cite{Lov1992}, obtained from a shallower survey, and highlights the need for further studies of the faint-end slope, at least down to magnitudes ${\rm M \sim -12}$.  

We believe that we have a fair knowledge of the local field LF, even though future surveys will provide better details, especially at faint luminosities.
Naturally, the number of low surface-brightness galaxies (LSB) we miss and the number of compact galaxies that could be confused with stars are still unknown.\\ 
According to \cite{Cal1987}, LSB galaxies are destroyed near the cluster center more likely than normal galaxies, and form a diffuse stellar background following the cluster potential. The compact galaxies, thanks to their strong density gradient, will survive also if their number is very high in a cluster, as found by \cite{Chi1972}, \cite{Ga2002} and \cite{Sak2002}.
Undoubtedly, the analysis of the complete SDSS will give an important step forward in solving these problems.

For clusters of galaxies the situation is more complex. A milestone in the study of the LF is the catalogue by \cite{Bi1985a} and the related analysis by \cite{Sa1985}. For the first time, they understood the role that the different types of galaxies play in the LF. They found that the faint end, clearly dominated by dwarf galaxies, had a rather high slope of  ${\rm \alpha \sim -1.4}$.\\
Because of the closeness of the Virgo cluster, these works should be taken as a firm reference point, and every study of the LF of cluster galaxies must be aware of the difficulties in  detecting and measuring LSB galaxies (\cite{Imp1997}).

A large amount of work has been published on the LF of cluster galaxies and we will discuss the recent literature in a comparative way in a forthcoming paper. To give a short survey of the situation, we refer here to a few works.

Studying A 963 (${\rm z \sim 0.2}$), \cite{Dri1994} find a result that coincides very well with the early work on Coma (${\rm z \sim 0.02}$) by \cite{GP1977}, and with the LF function derived for A 2554 (${\rm z \sim 0.01}$, \cite{Sm1997}). Their analysis on the sample, that reaches ${\rm M_{R} = -16.5}$, shows a LF that could be fit by a composition of two Schechter functions, or by a Schechter plus a power law, with slope $-1$ and $-1.8$ respectively (in reasonable agreement with the findings from the Virgo cluster and by the ESP in the field).
Also, this work outlines the uncertanties related to the background subtraction (Driver et al. (1994 Fig. 5a); see also \cite{Be1995}). As also shown by \cite{Ab1989}, the background density of galaxies is uncertain by a factor of the order of the effect we measure and shows changes from region to region. From published studies and from our own work, this seems to be the main source of bias in determining the slope of the LF faint end.

To better distinguish between the probable cluster members and the background, \cite{Biv1995}, used the redshift data in their work on Coma. Obviously, this would be the way to go, but generally redshifts are not available for the faintest objects. The spectroscopic sample used by Biviano et al., reaches magnitude ${\rm M_{b} = -16.9}$ (assuming for Coma ${\rm m-M=34.9}$ with ${\rm Ho = 75 km/s/Mpc}$), with a completeness of about 95\%. To extend the analysis to fainter magnitudes, ${\rm M \sim 15.5}$, they added 205 galaxies selected by photometric criteria and the faint end slope they measure is about ${\rm \alpha \sim -1.3}$.

Further confirmation of those results was obtained by \cite{Be1995}, who used the deepest observations available of the Coma cluster (down to ${\rm M_{R} = -9.4}$). They were not able to measure a detailed LF on the bright end, but found a slope ${\rm \alpha = -1.42 \pm 0.05}$ in the magnitude range ${\rm -19.4 < M_{R} < -11.4}$, and a sharp steepening of the LF beyond that range. This probably recalls the old problem posed by Zwicky (1972 - private communication) about the faint end of the LF: where does it end and what kind of objects populate it.\\ At faint magnitudes, it's indeed hard to distinguish the faint-magnitude objects of the cluster from background objects without high resolution imaging and spectra. 

Another interesting result has been pointed out by \cite{Bar2002}. They analyzed a very large sample of cluster galaxies with the same purpose as ours, that is to determine if the faint-end slope is a function of the cluster morphology and if there is a gradient in the faint end slope of the LF moving from the central cluster regions toward the outskirt. They seem to have reached some positive evidence on those effects even if they do not go very deep in magnitudes (${\rm M_{R} < -16}$). 
Finally, analysing the LF of the first cluster of our sample, A496, \cite{Mol1998}, found a clear evidence of bimodal LF and a quite high faint end slope (${\rm \alpha = -1.65}$). 
 
%----------------------------------------------------------------------------

\section{The luminosity Function}

The detailed description of the observations, the data analysis and the algorithms we used can be found in the papers by \cite{Mol1998}, \cite{Mor1999}, and the laurea thesis by \cite{Mor1997}, \cite{Ra1998}, and \cite{Paro2002}.
As we want to present the results from the analysis of 7 clusters of the original sample, we give here just few informations to ease the reading (see Table \ref{tab:7clu}).

\begin{table*}[!ht]
\centering
\begin{tabular}{|l|c|c|c|c|c|c|} \hline
Cluster      & ${\rm \alpha_{2000}}$ & ${\rm \delta_{2000}}$ & Redshift & B.M. type & Abell type \\ \hline\hline
Abell 0085   & 00h41m50.11s & -09d18m17.5s & 0.05560 & I    & 2 \\ \hline
Abell 0133   & 01h02m42.21s & -21d52m43.5s & 0.05660 & I    & 2 \\ \hline
EXO 0422-086 & 04h25m51.02s & -08d33m38.5s & 0.03971 & I-II & - \\ \hline
Abell 3667   & 20h12m35.08s & -56d50m30.5s & 0.05560 & II   & 2 \\ \hline
Abell 3695   & 20h34m46.86s & -35d49m07.5s & 0.08930 & I    & 2 \\ \hline
Abell 4038   & 23h47m41.78s & -28d08m26.5s & 0.02920 & I-II & 2 \\ \hline
Abell 4059   & 23h57m00.02s & -34d45m24.5s & 0.04600 & I    & 1 \\ \hline
\end{tabular}
\caption{The 7 clusters analized in this paper.}
\label{tab:7clu}
\end{table*}

We carried out the observations at ESO, La Silla, at the Danish 1.54 m Telescope using the Danish Faint Object Spectrograph and Camera (DFOSC) and the \textit{g}, \textit{r} and \textit{i} colors of the Gunn's photometric system (\cite{TG1976}; \cite{W1979}).

After a standard reduction of the data, for each cluster we created a catalogue of sources by merging the photometrical catalogue of the bright and extensive galaxies obtained through a program specifically developed by Moretti et al.(1999) with the photometrical catalogue of small and faint objects obtained using the MIDAS Inventory package.
Infact, as Inventory was developed to detect sources in distant clusters (point-like sources), we first extracted and analysed all the bright and extended galaxies through the program by Moretti et al. (1999), that reproduces the image of a galaxy through the Fourier analysis of its isophotes and subtracts it to the main frame, then applied Inventory.\\
The catalogs thus created were tested through a series of simulations, briefly described below, to evaluate their completeness:
using one of the galaxies reproduced with the program by Moretti et al. divided by coefficients ad hoc, we created a sample of artificial galaxies with a known distribution of magnitudes. We partitioned the field to be tested in circular annuli in which we added the artificial sample with known random coordinates; by running Inventory on the added fields we were able to estimate the selection function as a function of the distance of the cluster center.

After that, we carried out our analysis.

\subsection{The standard analysis}

\begin{figure}[!ht]
\centering
\includegraphics[width=8.5cm]{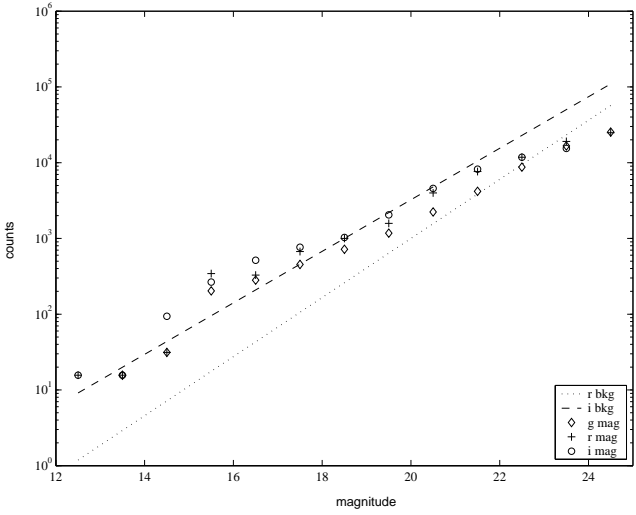}
\caption{A0085. The total counts over the whole area observed are plotted together with the background counts as measured by \cite{Ty1992}. Due to the large background component on the outermost area for the faintest galaxies, the counts on the cluster galaxies almost coincide with the background counts.}
\label{fig:1}
\end{figure}

To sharpen the contrast between the counts in the cluster and the counts in the field, we selected those galaxies that in the color magnitude plot are placed in between the 68\% confidence curves defined by the line fitting the \textit{r} versus \textit{(g-r)} distribution of bright galaxies (the equivalent of the E/S0 sequence as defined by \cite{VS1977}) and its rms (Fig. \ref{fig:1}). These galaxies, mostly ellipticals, have indeed a higher probability to be cluster members.

This area is defined by the curves:
\begin{equation}
{\rm C = a \cdot m + q \pm \left(k\cdot \sigma_{fit}+\sigma_{0}\right)}
\label{eq:cmr}
\end{equation}                              
Where the term ${\rm a \cdot m + q}$  represents the fit of the color-magnitude relation for galaxies brighter than ${\rm r = 19}$ and  ${\rm \sigma_{0}=\sqrt(\sigma_{g}^{2}+\sigma_{r}^{2})}$ is the color index error derived using the magnitude error estimated on the fields overlapping in the different frames.\\ 
Non-members may perturb the fit to derive the coefficients $a$ and $q$, also in view of the small number of stars which belongs to the bright sequence, so we coupled the regression solution with the best fit as estimated by eye. 

We assumed that all the galaxies which in the color-magnitude diagram belong to this region form what we will define \textit{sequence galaxies} and charachterize the cluster sample. These are the objects we used to estimate the parameters of the distribution function by the method of the maximum likelihood.\\
\begin{table}[!h]
\centering
\begin{tabular}{|c|c|c|c|c|}\hline
Cluster & All & Sequence & Sequence         & ${\rm R_{max}}$ \\
        &     &          & (within $400''$) &    (arcsec)     \\\hline           
A0085   & 2701 & 1688 & 659 & 1473.2\\
A0133   & 4022 & 3068 & 1016 & 1350.6\\
A3667   & 3205 & 1887 & 521 & 1597.9\\
A3695   & 2156 & 1429 & 751 & 1293.1\\
A4038   & 3955 & 2949 & 767 & 1714.3\\
A4059   & 1778 & 1201 & 543 & 1107.0\\
EXO 0422-086 & 1612 & 1254 & 604 & 1157.8\\\hline
\end{tabular}
\caption{Number of galaxies for each cluster and maximum distance from the cluster center covered by the data (${\rm R_{max}}$). The clusters have been observed up to an everage distance of about 23 arcmin or 1.08 Mpc, that is the observations cover almost an Abell radius and therefore we observed up to a distance from the center of the order of ${\rm r_{200}}$.}
\label{tab:numgal}
\end{table}

We assumed that the sequence galaxies are, as stated above, in large part cluster galaxies. This is true especially for the brighter objects where the background is pratically absent. In other words, we subtracted only those objects outside the sequence where it is more likely that the background dominates at faint magnitudes.\\ 
While this assumptions has no effect in the estimation of the Luminosity of the clusters since that is due essentially only to the bright cluster objects, it may seriously affect the estimate of the slope of the faint end of the LF. Indeed instead of measuring the slope of the faint end of the Luminosity Function we could be measuring the slope of the background counts.
To demonstrate that by selecting the sequence galaxies as the sample on which to measure the LF parameters we do not bias the background contamination, we estimate, for a few particularly rich clusters, the Luminosity function after background subtraction and obtained values in perfect agreement with what we obtained by the analysis of the sequence galaxies and the use of the maximum likelihood algorithm.

The background counts of galaxies in the Gunn and Thuan photometric system were estimated using the counts published by \cite{Ty1992}.

The background counts as measured by \cite{Ty1992} were used instead of the SDSS counts since, at the time we started the observations and the analysis these were, at the best of our knowledge, the best counts we could use at different wavelengths. Furthermore, these counts were in good agreement with the counts measured in our fields using the outermost fields observed in those clusters where we estimated the background. Moreover the SDSS counts cover a smaller dynamical range and stops at ${\rm m\sim20}$ and the early attempts we made, using preliminary SDSS data as distributed in the Web, to extrapolate to fainter magnitude the slope (down to about ${\rm r\sim 25}$) showed that we were overestimating the background counts.

Indeed by plotting the coefficients of the logarithmic fits of the counts given by Tyson as a function of the effective wavelength of the pass-band filter, we notice that these observed values are accurately fitted by a polynomial function. It is then straightforward to estimate by interpolation the parameters expected for counts in the Gunn and Thuan photometric system, by so doing we obtain the following relations:

\begin{equation}
\begin{array}{lcccr}
{\rm \log{N_{g}}=(0.45\pm0.056)g-(6.30\pm0.89)}\\
{\rm \log{N_{r}}=(0.39\pm0.063)r-(4.67\pm1.01)}\\
{\rm \log{N_{i}}=(0.34\pm0.070)i-(3.38\pm1.11)}\\
\end{array}
\label{eq:bkg}
\end{equation}

The errors given in the above relations represent the progressive errors estimated on the interpolation relation by propagating the errors of the fits relating the parameters to the effective wavelength of the filters used by Tyson.

\begin{figure}[!h]
\centering
\includegraphics[width=8.5cm]{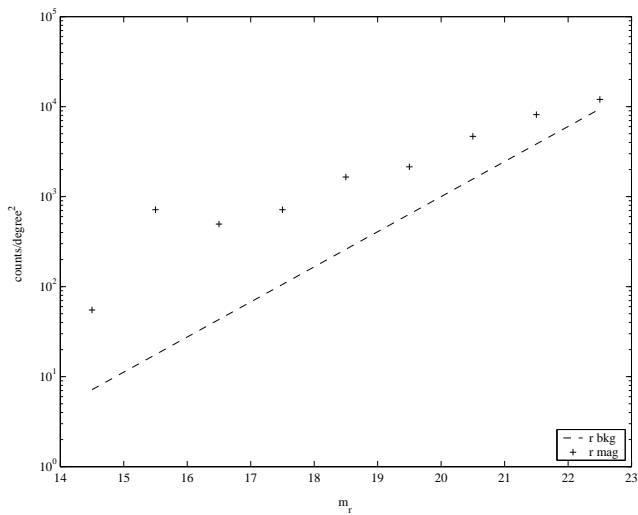}
\caption{A0085. Counts in the r filter limited to the central frame compared to the background counts as given by \cite{Ty1992}. The cluster signal is well visible over the background.}
\label{fig:2}
\end{figure}

We modeled the distribution in luminosity by using a bivariate LF that is the sum of a Gaussian and a Schechter. The choice is the most reasonable one following the work by \cite{Sa1985} on the Virgo cluster; it is evident from the data of the present analysis, Fig. \ref{fig:1} and Fig.\ref{fig:2} and as indicated also in \cite{Mol1998} and references therein. In a cluster, as it is very evident also from our data, the bright galaxies are located in the central region and these are the galaxies that are fitted by the Gaussian component of the LF. That is the two component fit is valid only for the central region, in the outskirts of the cluster the luminosity of the galaxies is distributed according only to the Schechter function (Fig. \ref{fig:add}).

We are aware that not all of the published analysis on the LF of clusters of galaxies, such as the one of \cite{Got2002}, evidence the bimodal distribution. While it is hard to answer why without remaking the analysis and eventually the data reduction, it seems to us that the main reasons reside on the statistics, accuracy of the magnitudes and background subtraction and, in part, from the binning. Indeed from a practical point of view, and looking at it in a slightly different way, the main evidence depends from the small ``\textit{gaps}'' in the observed distribution which is evidenced in the point where the composite functions are both somewhat weaker. In our cluster sample this occurs at the \textit{r} mag which is in the range $17 - 18$. Obviously in a redshift sample the signal vs. noise ratio is large due to a better selection of cluster members detaching the cluster galaxies almost completely from the background. Indeed this is clearly shown also by the Coma sample selected by \cite{Biv1995} and by the sample in \cite{Mor1999}. On the other hand the analysis of \cite{Got2002} based on the SDSS survey clearly shows that a) the analysis is consistent with the distribution of two underlying populations, b) the analysis deals with the composite LF so that details are smoothed out adding the contribution of different clusters and c) the limiting magnitude is not faint enough (the sample stops at ${\rm M\sim -18}$) to clearly define the faint end.

For the Maximum Likelihood fitting we used only objects brighter than the 14th magnitude. This means that we excluded the cD (in two clusters this cut off excludes 2 or 3 galaxies) from the fitting. The cD does not necessarily conform to a distribution function as the other galaxies and its luminosity may be a function not only of the cluster mass distribution at the time of formation but, above all, may reflect the characteristics of evolution. While it is easy to account for its luminosity, it is a perturbing factor in the fitting. The cD galaxies will be discussed in detail in a forthcoming paper.

We analyzed all the clusters in a similar way to have internal consistency. In most cases, two of the authors analyzed the clusters and compared the results to have an estimate of how much the result could depend from our method of analysis and have, at the same time, an estimate for the uncertainties. The results of the standard analysis are listed in Tables \ref{tab:riski1} and \ref{tab:riski2}. Using these values and the master photometric catalogues we derived the Luminosities listed in Table \ref{tab:lums}.

\section{Checking our analysis, a few test cases}

In this section, we illustrate part of the data and our analysis also evidencing some weak points. While the standard analysis was done in a very homogeneous and systematic way, for most of the clusters we used alternative analysis, that is a) we estimated the luminosity function subtracting from the central frame the counts as given by \cite{Ty1992}; b) we subtracted the background using the outermost field we observed where the cluster galaxies contamination is known to be negligible.
In all these cases, after the subtraction of the background, the LF parameters were estimated after binning the derived histogram, assumed to represent statistically the counts of the cluster galaxies, and using a ${\rm \chi^{2}}$ minimizaton program.

As it is explained below, it makes a difference whether or not the cD galaxy is part of the sample. Furthermore, and as expected, if the analysis is not limited to the central region, of the order of a core radius, the contrast cluster vs. background is largely washed out. But, as described in the following text, the best evidence that the bimodal distribution is a real fact is illustrated in the lower panel of Fig. \ref{fig:5}. Here the contamination of the background galaxies brighter than the ${\rm 21^{st}}$ magnitude is negligible and the maximum of the Gaussian distribution is readily visible at ${\rm r\sim16}$ while it is also clearly visible the bright end of the LF and to the low number of galaxies expected toward the faint end of the Gaussian distribution.

Since all the set of clusters used to test the method were treated in a similar way, for each cluster we discuss only part of it in order to give a complete view of the tests we made in oder to develop a feeling for the robustness of our results. By doing so we also avoid useless duplications. We also definitely think that the standard analysis and the results we give in Tables \ref{tab:riski1} and \ref{tab:riski2} represent the consistent results, with their errors, that are the output of this work and that is why we prefer not to confuse the issue by listing the derivation of the parameters by other methods that while in agreement within errors with the standard analysis are less homogeneous and rigorous.

\subsection{A0085}

In Fig. \ref{fig:1} we plot the total counts, that is the counts made over the whole observed area of our sample, for the 3 colors used (\textit{g}, \textit{r} and \textit{i}), as a function of the magnitudes. Moreover, we plot the counts for the field as derived using the analysis by \cite{Ty1992}, in the filter \textit{r} and \textit{i}.\\ Our counts do not differ much from the background from one filter to the other. The larger discrepancy is with the \textit{i} filter in which we count more than 30\% of galaxies less than in the other filters. A probable cause for this is the high sky background and the lower quality of these frames. However, we want to evidence the excess of counts (bump) compared to the background at magnitude $\sim 15$ that is due to the cluster we observed.\\ 
If we plot only the central field, the contrast cluster/background incerases considerably and it evidences indeed the fact that the cluster galaxy counts can be clearly separated. This is naturally what we expect. Assuming a King's density profile the percentage of cluster galaxies we would expect on the outermost field we observed in the different clusters is of about 10\% so that here the backgrounds dominates. Indeed, and always to check out procedures, we also estimated the cluster galaxies counts by subtracting the outermost field from the central once assuming that the small cluster contamination on the background so defined would not bias the estimate od the cluster LF. We got a good agreement, for those clusters which were analysed also in this way, with the standard analysis.
 
The bright end is most highly dominated by cluster galaxies as we expect a small number of field galaxies in such a limited volume of space.

\begin{figure}[!ht]
\centering
\includegraphics[width=8.5cm]{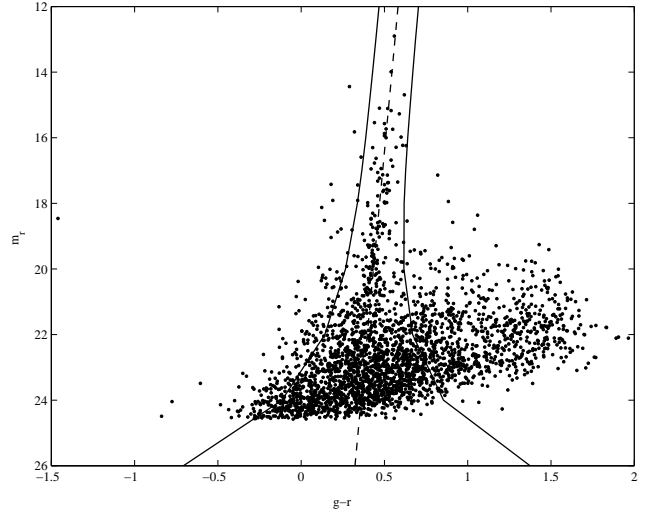}
\caption{The color magnitude plot for the central area of the cluster A0085. The plot illustrates the fit, shown here as a dashed black line, that determines the sequence galaxies as described in the text. The solid lines, as described in the text, define 68\% probability that the objects has a color-magnitude as defined by the sequence.}
\label{fig:3}
\end{figure}

After the correction for completeness, it is straightforward to estimate the cluster contribution by subtracting the field counts. However, by doing so, we are more sensitive to the fluctuations of the background and to an eventual uncontrolled incompleteness introduced by our analysis. Furthermore, at the bright end we are dealing with a few objects, and the low statistics make things more uncertain. That is why we decided to use only relative measurements and, to further increase the contrast cluster versus field, to use only a selected subsample of galaxies for each cluster delimited by the fitting of the bright sequence galaxies and the ${\rm 1\sigma}$ error.

As we explained earlier, we define a color-magnitude sequence via the color magnitude diagram for each filter and for each cluster. In the case of the \textit{r} observations for the cluster A0085, Fig. \ref{fig:1}, the lines delimiting the sequence galaxies area are defined by: 
\begin{displaymath}
\begin{array}{l}
{\rm (g-r)_{seq}=-0.0180\cdot r + 0.802}\\
{\rm \sigma_{err-cur}= (g-r)_{seq}\pm \left(0.117+0.675\cdot0.217\right)}\\ 
\end{array}
\end{displaymath}
For this cluster, we used a factor 0.675 to reduce to 50\% the probability that a galaxy belongs to the sequence with the computed rms. We also extended the sequence multiplying by a factor 3 the value of ${\rm \sigma_{fit}}$ to increase the number of sources at the faint end and to ease the comparison with the other clusters (in particular A4059). These minor adjustments do not affect the derivations of the parameter in which we are interested. On the other hand, they are selected during the analysis to optimize the statistics and the homogeneity of the solution. Two members of our group derived the luminosity function both using the sequence we just defined and all of the central field objects. This checking analysis was also carried out using the ${\rm \chi^{2}}$ method and subtracting the background. For the faint end of the LF, the resulting slopes were all very similar to that obtained with the standard analysis. In all cases, we obtained similar and consistent results within the errors, an evidence that the results are good and practically unaffected by the method of the analysis.

\begin{figure}[!ht]
\centering
\includegraphics[width=8.5cm]{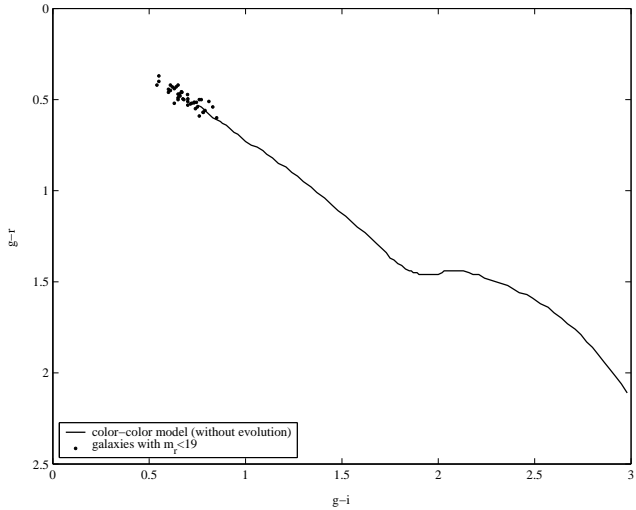}
\caption{Color-color plot of A0085. The bright sequence galaxies lays exactly on the color-color curve generated by the models of synthesis.  Naturally, the sub-sample sequence fits as well with a broader dispersion and reflects the color-color cuts we imposed in the definition of the sequence.}
\label{fig:4}
\end{figure}

\subsection{A0133}

With this cluster we had some difficulties in getting a reasonable fit of the observed LF in spite of the large number of measurements. As we did for most of the clusters, we carried out the fits in the different wavelengths deriving the parameters of the LF:
\begin{itemize}
\item[a)]  for all the galaxies in the central field after subtracting the background;
\item[b)]  for the sequence galaxies of the cluster after background subtraction;
\item[c)]  for the counts derived as the difference between the central field and the outermost field (for the clusters for which we could use the latter as a background).
\end{itemize}
In spite of the large uncertainties, we derived the same parameters within errors with all methods. Finally, the various samples were tested for solution in a semi-empirical way. The counts as a function of the magnitudes, were fitted by modifying the parameters derived statistically and estimating the best fit, or the different fits, also by eye.  This procedure, that we describe only for A0133, was indeed carried out for most of the sample clusters. The reason is that in some cases the parameters we derived formally using the Maximum Likelihood have large errors. This is especially true for the magnitude of the knee of the Schechter function or for the ratio of the Gauss to the Schechter normalization factor. By doing so we hoped to avoid flukes due to the analysis and to the small counts and to make sure which parameters must be taken with caution. What remains certain, however, is that the LF is in all cases dominated by a rather broad Gaussian defined by the brightest galaxies and dominating over the Schechter function which is essentially defined by the fainter galaxies. The cD galaxy does not fit the Luminosity Function.

\subsection{A4038}

\begin{figure}[!ht]
\hfill
\centering
\begin{tabular}{c}
\includegraphics[width=8.5cm]{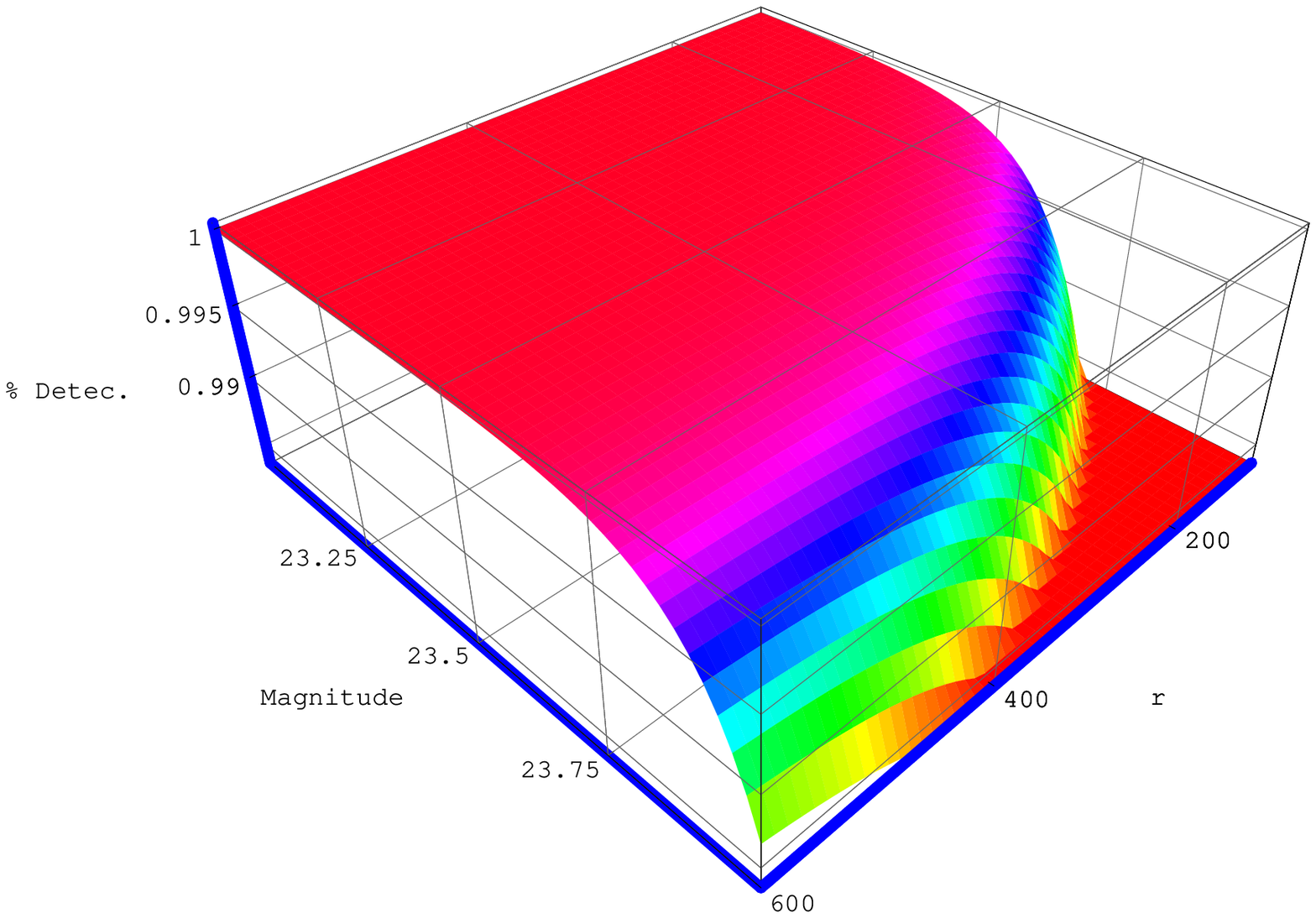}\\
\includegraphics[width=8.5cm]{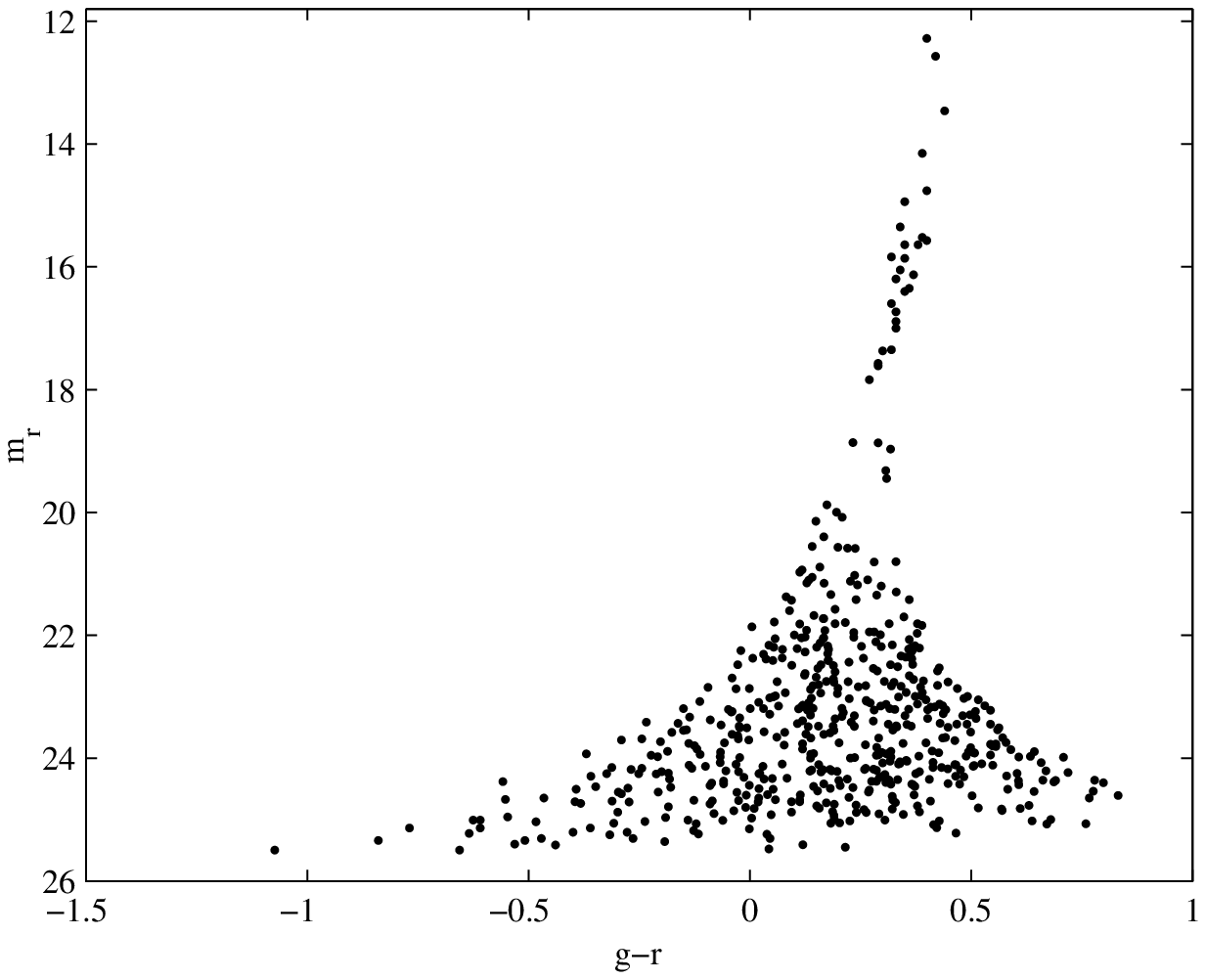}\\
\end{tabular}
\caption{Abell 4038. Upper panel: the completeness Function for the \textit{r} filter as a function of the distance from the cluster center (in seconds of arc). Lower panel: the distribution of the sequence galaxies shows that the number density, and therefore the distribution function we derived, is composed by a condensation of bright objects and by a multitude of faints objects whose number increases with the magnitude. }
\label{fig:5}
\end{figure}

With a completeness down to the 22nd magnitude in the central region and down to the 23rd magnitude in the outskirts (upper panel of Fig. \ref{fig:5}), A4038 is a clear example of the bimodal distribution function.

The lower panel of Fig. \ref{fig:5} shows the color-magnitude plot of the sequence galaxies. We see quite clearly the increase first and the decrease later of the number density of galaxies along the sequence going from brighter to fainter magnitudes, and this effect has been observed in all clusters and in all the colors. At about ${\rm r = 19}$, we find the so called \textit{gap}, that is the region in the LF where the Gauss distribution dominating at bright magnitudes merges with, and is taken over, by the Schechter distribution.\\
In the outermost field of this cluster the background, as estimated by \cite{Ty1992}, is somewhat higher than our counts for ${\rm r \ge 21.5}$. This outlines that at faint magnitudes we are strongly affected by the fluctuations of the background and by incompleteness.

\subsection{A4059} \label{sec:4059}

\begin{figure}[!ht]
\centering
\includegraphics[width=8.5cm]{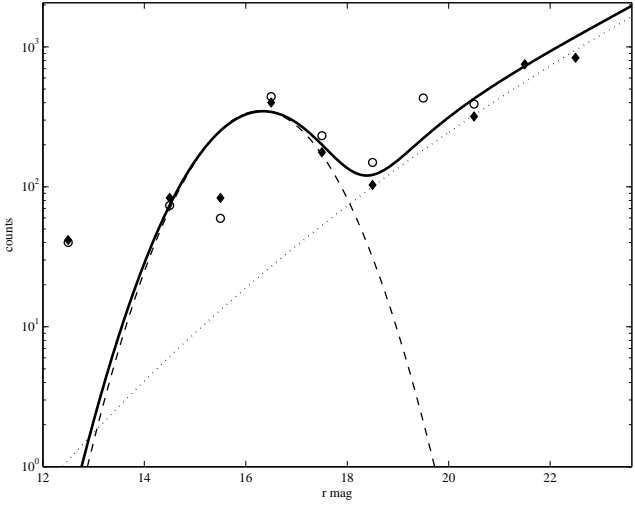}
\caption{A4059. The black filled diamonds refer to the counts in the central 400$"$ frame after subtracting the counts by \cite{Ty1992}, while the empty dots refer to the same counts after subtraction of the outermost field we observed for this cluster. The solid line reflects the values derived from a semi empirical fit of the data. The dash line show the Gaussian distribution selected and the fit of the faint end by a straight (dot) line. The fit has been obtained using the following values for the LF parameters: ${\rm G/S=350.0/150.0}$, ${\rm m_{G}=16.3}$, ${\rm \sigma_{G}=1.0}$, ${\rm m_{S}=18.0}$ and ${\rm \alpha=-1.5}$. The cD galaxy does not fit the LF.}
\label{fig:7}
\end{figure}

We use of this cluster to discuss in more detail the fit and related uncertainties on the study of the LF of clusters. These uncertainties are not due to pure statistics, but rather to the selected procedures. As usual, we illustrate the \textit{r} filter observations for uniformity with the previous, and following,discussion. A similar analysis has been done and checked in all colors on various clusters obtaining consistent results. That is to say that the goodness of the parameters values obtained and the procedures discussed is robust. 

In Fig. \ref{fig:7}, we plot both the data obtained by subtracting the background as estimated by \cite{Ty1992} and that derived by our data using the outermost field observed (as we stated earlier here we have a cluster galaxies contamination of about 50\%). The Tyson Background is somewhat higher, the difference being probably due to fluctuations. Contrary to what has been done for standard solution, in the distribution in Fig. \ref{fig:7} we also included the cD galaxy that, being too bright, does not fit the LF, no matter which parameters we select. 

At the faint end we obtain a fit of the counts that is in excellent agreement with the value obtained using the Maximum Likelihood method applied to the sequence galaxies. This justifies the method once more and underlines that we derive reliable values also for the faint end in spite of correcting only partially for the background galaxies. 

However, to stress the subtleness of the faint end fitting in presence of a background contamination, we point out that this slope can easily be estimated by a linear fit of the faint galaxies. The faint end of the Schechter function expressed in magnitude and on a logarithmic scale is given by the relation ${\rm -0.4\cdot(m-m^{*})\cdot(\alpha+1)}$ so that ${\rm \alpha}$ can be derived by a simple linear fit of the faint end of the observed distribution using the relation: ${\rm \alpha=-\left(1+(\textit{observed slope}/0.4)\right)}$. The background can also readily checked and subtracted since we simply make the difference between two straight lines. Assuming the counts are dominated by background objects, we would have ${\rm \textit{observed slope} = 0.39}$ and derive ${\rm \alpha > -1.5}$ and closer to ${\rm -2}$.

\section{The cluster luminosity: the contributions of the bright and the faint end}

An estimate of the luminosity of a cluster is given by simply integrating the Cluster LF, in absolute magnitudes, using the following expressions:
\begin{equation}
\begin{array}{l}
{\rm L_{S}=-1.086\cdot \int^{\frac{L_{2}}{L^{*}}}_{\frac{L_{1}}{L^{*}}} \left(\frac{L}{L^{*}}\right)^{\alpha +1}\cdot e^{\frac{-L}{L^{*}}} d\left(\frac{L}{L^{*}}\right)}\\
{\rm L_{G}=-2.5\cdot\int_{\log{L_{1}}}^{\log{L_{2}}}L\cdot\rho\cdot e^{-0.5\cdot\left(\frac{-2.5\left(\log{L}-\log{L_{G}}\right)}{\sigma_{G}}\right)^{2}}d(\log{L})}\\
\end{array}
\end{equation}
where ${\rm L_{1} >L_{2}}$ and where ${rm \rho}$ is the Gaussian/Schechter normalization ratio.

As we said before, in the expression of the LF the faint end is the greater source of uncertainties because of the background fluctuations and of the uncertainties in the estimate of the completeness (the estimate of the selection function remains often uncertain in spite of the Monte Carlo approach which is less robust when applied near the core area where many bright galaxies are located).

\begin{figure}[!ht]
\centering
\includegraphics[width=8.5cm]{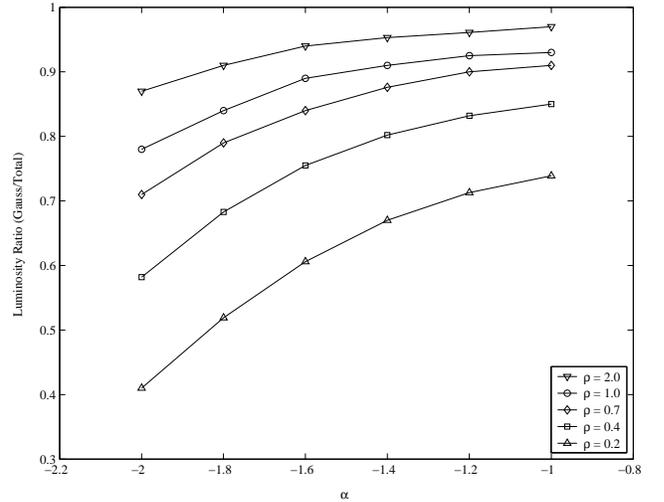}
\caption{The Ratio between the luminosity contained in bright galaxies fitted by a Gaussian LF and the total luminosity (Gaussian + Schechter function) as a function of the parameter ${\rm \alpha}$ and the normalization ratio between the Gaussian and the Schechter function, ${\rm \rho}$ in the equation. From top to bottom ${\rm \rho=2.0, 1.0, 0.7, 0.4, 0.2}$ .}
\label{fig:8}
\end{figure}

It seems that the cluster luminosity is dominated by the bright galaxies. In Fig. \ref{fig:8} for various values of the ratio between the Gaussian and the Schechter normalization, we plot the percentage of the total cluster luminosity which is due to the bright galaxies as fitted by a Gaussian. We used a magnitude difference ${\rm <M_{Gauss}> - M^{*}_{Schechter} = 1.5}$. This difference is close to what we find in some of our clusters. In the Virgo cluster, \cite{Sa1985}, find ${\rm <M_{Gauss}> - M^{*}_{Schechter} = 0.9}$. However, the ratio is practically not affected very much by such differences. As it can be seen from Fig. \ref{fig:8}, what counts is mainly the normalization ratio and the value of the faint end slope ${\rm \alpha}$. Disregarding extreme values, that is ${\rm \rho>1}$, which have been however also observed in our clusters, we find that for ${\rm \alpha \sim -1.4}$ the ratio varies between ${\rm 0.6}$ and ${\rm 0.85}$ while for ${\rm \alpha \sim -1}$ the ratio changes between ${\rm 0.74}$ and ${\rm 0.92}$. Clearly, in those cases where ${\rm \rho > 1}$, the Gaussian distribution function contribution dominates. In the Virgo cluster (Sandage et al. 1985) we estimate ${\rm \rho \sim 0.14}$. However, in this cluster ${\rm \sigma_{G}}$ is rather large while ${\rm \alpha \sim -1.35}$ close to what has been found in the majority of the clusters.

Merging clusters, or more precisely clusters for which substructures are detected, do not affect the estimate of the total luminosity and the LF. During the merging of substructures we expect changes in luminosity of the galaxies due to induced tides and therefore luminosities for the merging galaxies that differ somewhat for the luminosities of the merged galaxies. On the other hand this effect, which likely result in a loss of stars to the Intra-Cluster Medium (ICM) and at the same time brightening and star formation due to the induced turbolence in the Interstellar Medium (ISM), is a secondary, not detectable effect to the analysis of this work.

We calculated that an estimate of the cluster luminosity using only the bright galaxies as fitted by a Gaussian plus a standard slope faint end luminosity function gives a mean error of about 10\%. In all those cases in which there are uncertainties in the faint end fitting or we notice the presence of a greater background contamination, we estimated, as a check, also the total luminosity by using mean parameters for the faint end of the LF.

In general, the procedure we followed is the following:

\begin{itemize}
\item[-] We used the cluster parameters as derived from the standard analysis
\item[-] We normalized the luminosity function by subtracting the luminosity derived by the counts of the background as estimated by \cite{Ty1992} from the counts we have for each cluster in a given range of magnitudes selected ad hoc (the range is not critical):  $g < 19.5$, $r <19$ and $i < 19$. 
\item[-] We limited the normalization to the bright galaxies in order to avoid further uncertainties due to the background and the subtraction of large numbers at faint magnitudes.
\item[-] In most cases, we based the normalization on the Gaussian component since the ratio of the Gaussian to Schechter was estimated from the fit, albeit with large error in some cases. For the normalization, we used only the frame about the cluster center.
\item[-] For each cluster we also counted and selected the galaxies in the range of apparent magnitude corresponding to $M_{g} < -16.5$, $M_{r} < -17.0$ and $M_{i} < -17.0$ and we simply added up the relative luminosities.
\end{itemize}

We observe a clear difference between the central part of the cluster, where the bright galaxies dominate, and the outskirts, where the central part of the cluster bright galaxies are generally absent. This is a strong evidence of segregation in luminosity and is consistent with the model of a fairly relaxed cluster of galaxies. That has been taken into account in computing the cluster luminosity.

\begin{figure}[!ht]
\centering
\begin{tabular}{c}
\includegraphics[width=8.5cm]{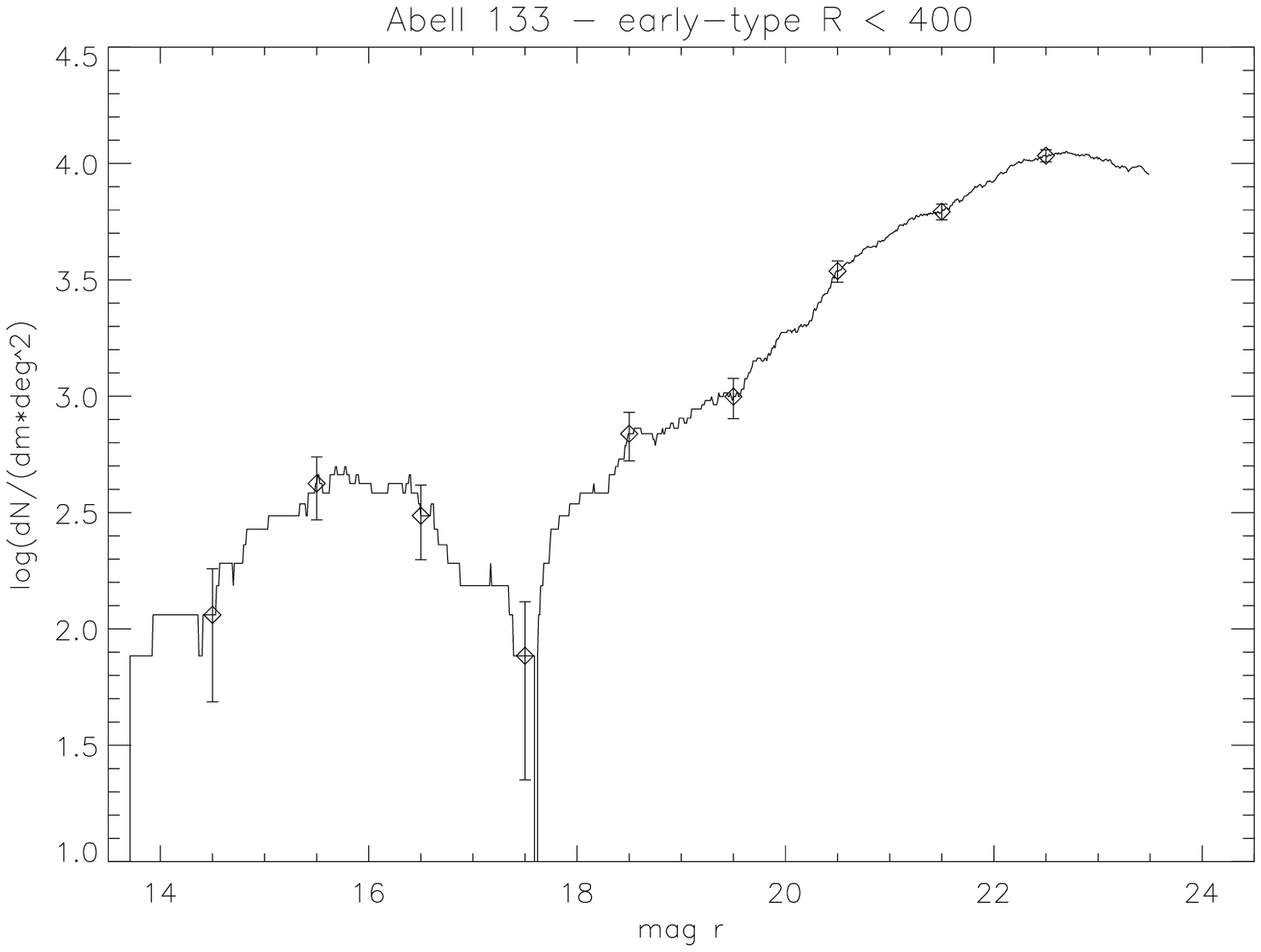}\\
\includegraphics[width=8.5cm]{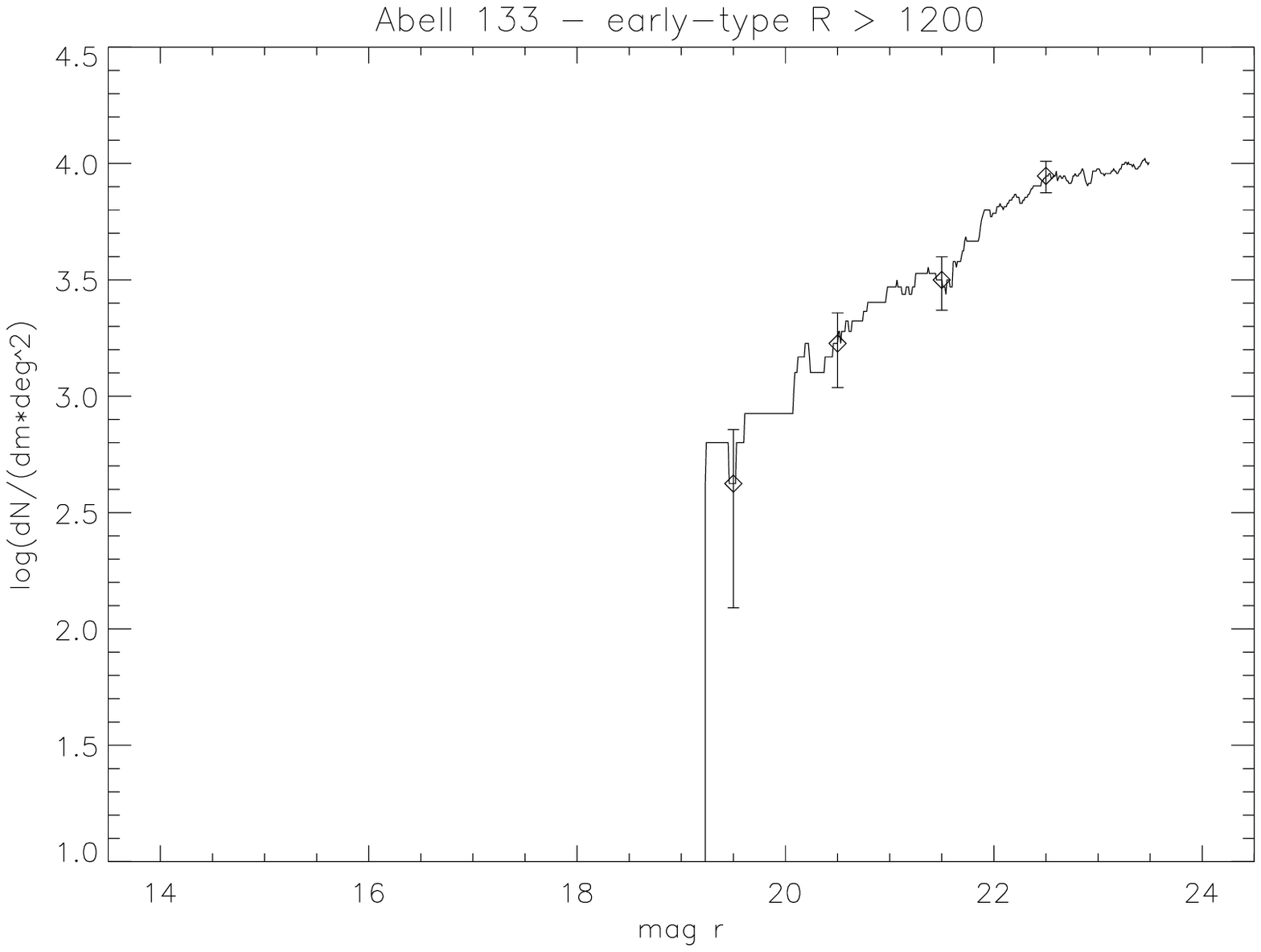}\\
\end{tabular}
\caption{A0133. For the sequence galaxies within 400$''$ from the cluster center the LF is a Gaussian plus a Schechter (upper panel), while for the sequence galaxies more distant than 1200$''$ from the cluster center (lower panel) the LF is well fitted by a Schechter function.}
\label{fig:add}
\end{figure}

To better account for this difference we adopted a model in which the number density distribution follows a King's profile and the LF changes from a Gaussian plus a Schechter (where the luminosity of the Gaussian is dominant) to a Schechter function from the central regions to the outskirt the cluster (Fig. \ref{fig:add}).

To estimate the contribution of the outskirt of the cluster to the luminosity, we thus assume a decreasing density obeying the King's law and account for the fact that we do not observe bright galaxies in the most distant regions of the cluster. We extrapolate the estimate to 1 Abell radius (${\rm AR = 1.5 h^{-1}}$ Mpc) and we use a core radius of ${\rm 0.25 h^{-1}}$ Mpc in all those cases for which we do not have a direct estimate of the core. The normalization of the King's function, ${\rm \phi^{*}(r)}$, is given by the relation:
\begin{equation}
\centering
\begin{array}{l}
{\rm \int^{r_{e}}_{0}\phi^{*}(r)\phi(M)drdM=\int_{Cl.Field}<\phi^{*}>\phi(M)dM}\\  
\end{array}
\end{equation}
with ${\rm r_{e}=\sqrt{\frac{Area(Cl.Field)}{\pi}}}$, where with "Cl. Field" we generally refer, unless otherwise stated, to the observed central frame.

As we have shown in the previous section, the cD galaxy, or the two brightest galaxies, hardly fit the analytical functions we used. Indeed depending on how the brightest galaxies form, or grow, it may be reasonable to expect a peculiar behavior and disagreement with the analytical function. In other words the analytical fit may underestimate the Luminosity because such a bright galaxy is not expected by the distribution. To compute the luminosity, the LF has been considered in the range between ${\rm M=-12}$ and ${\rm M=-24}$.
	
\subsubsection{A0085}

To have reasonable statistics and a small background contamination, we normalized the luminosity function with the counts of the cluster galaxies within a distance of 300" from the cluster center, that is about the area of the cluster core. The observed area was computed from the geometry of the observations and accounting in an approximate way for the occultation due to the cD and a few other bright galaxies (a rather small correction however). To avoid a large correction for incompleteness, we also limited the counts to galaxies brighter than ${\rm r = 23.0}$ (${\rm M_{r} =-14.61 + 5\log{h}}$), while for the background we used the counts as given by \cite{Ty1992}.

\begin{figure}[!ht]
\centering
\begin{tabular}{c}
\includegraphics[width=8.5cm]{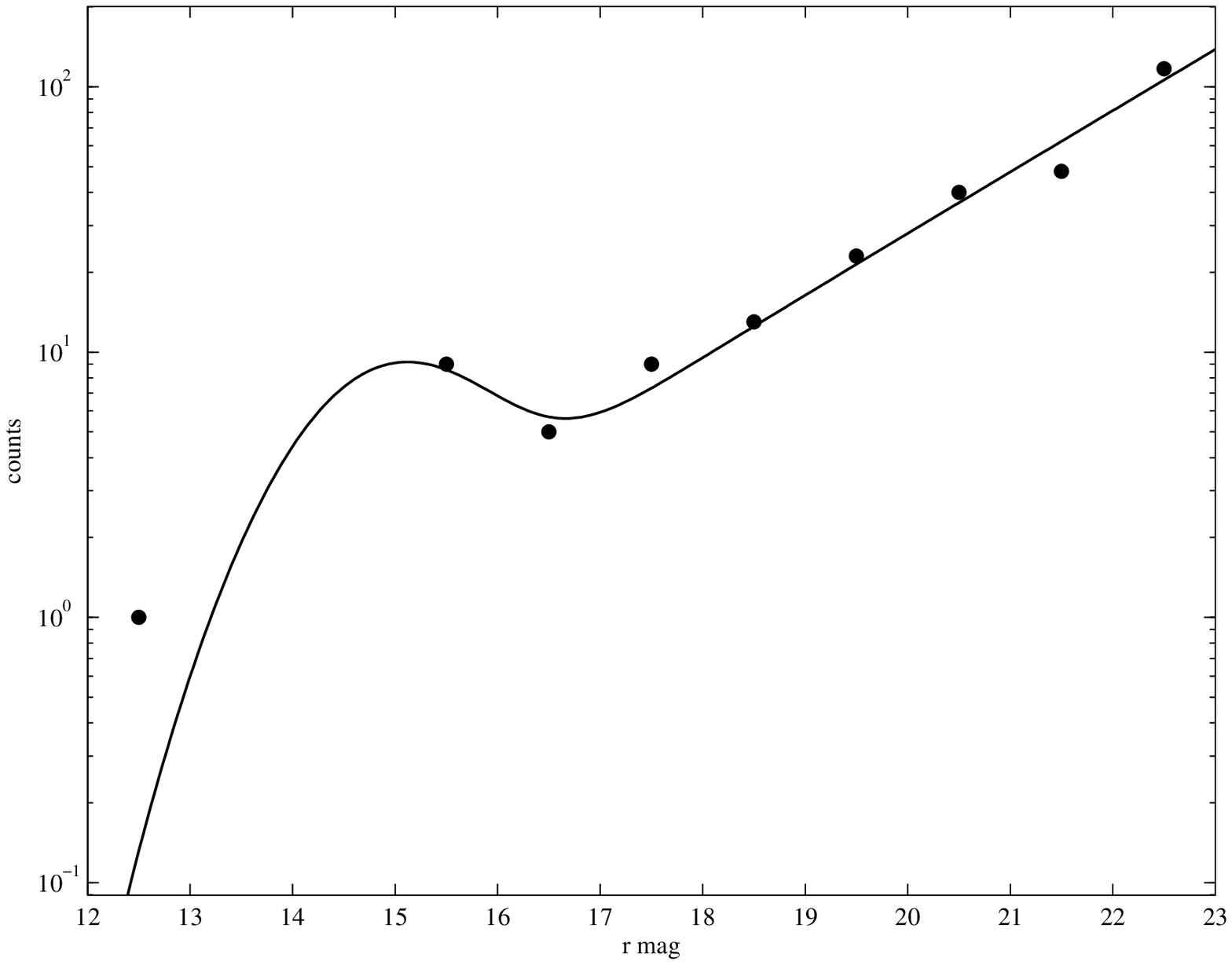}\\
\includegraphics[width=8.5cm]{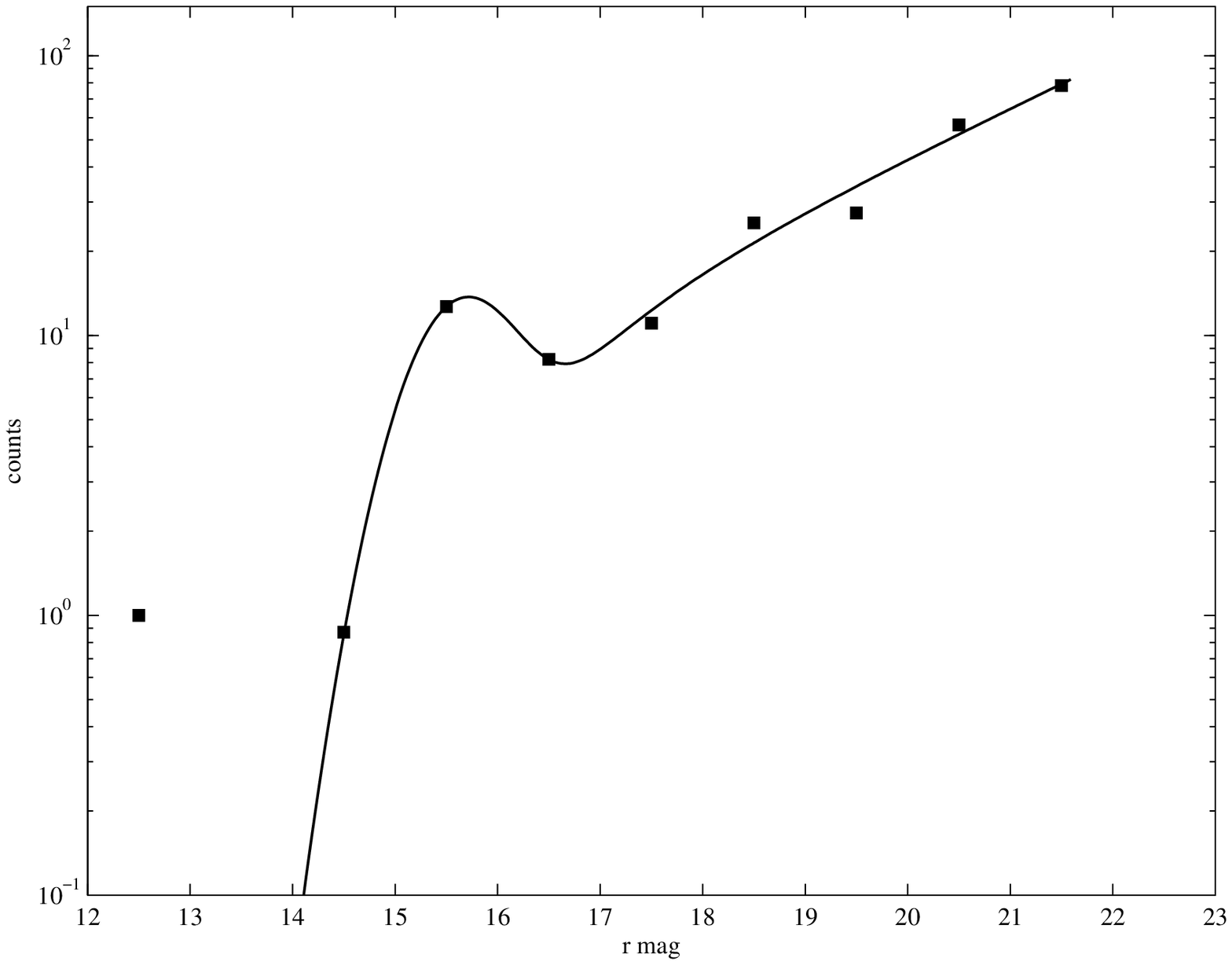}\\
\end{tabular}
\caption{Two possible fits of the LF for A0085: excluding the cD galaxy (upper panel), where the errors are simply the square root of the counts, and with a different binning trying to include with an optimization by eye the cD galaxy (lower panel). See however the appendix for the formal solution without cD.}
\label{fig:9}
\end{figure}

Normalizing, the luminosity distribution function using the observed galaxies within 300$"$, we compute a luminosity in the \textit{r} band of ${\rm L_{300''}= 9.1\cdot10^{10} L_{\odot}}$, with 46\% of this given by the bright Gaussian component. The contribution of the Schechter component within one AR is ${\rm L_{AR}=1.08\cdot10^{11} L_{\odot}}$, so that the total luminosity of the cluster (as we said there are no bright galaxies outside the core radius) within an Abell Radius is ${\rm L_{Tot AR} = 1.99\cdot10^{11} L_{\odot}}$. A somewhat larger value is obtained if we normalize the LF to the bright galaxies only. In this case, we derive ${\rm L_{300''} = 1.26\cdot 10^{11} L_{\odot}}$ and ${\rm L_{Tot AR} = 2.76\cdot 10^{11} L_{\odot}}$.

The parameters we derived using the standard solution for the LF, Tables \ref{tab:riski1} and \ref{tab:riski2}, are rather stable in the sense that were obtained, in agreement within errors, using different methods: ${\rm \chi^{2}}$ and Maximum likelihood. Unlike most of the other clusters we analysed, we find a value of ${\rm \rho}$ (the ratio between the Gaussian and the Schechter function coefficient) larger than one, a rather small ${\rm \sigma_{G}}$ and a magnitude for the knee of the Schechter function (${\rm m^{*}}$) which is brighter than the mean magnitude of the Gaussian distribution (${\rm m_{G}}$). This is suspicious and certainly due to the fluctuations related to the very few brightest galaxies. We can account of this by using a semi automatic (in part also by eye) fitting with the result, quite reasonable, of the lower panel of Fig. \ref{fig:9}, where the brightest galaxy has also been plotted.

\subsubsection{A0133}

For A0133 we measure a rather steep faint slope. However, we may be somewhat contaminated by background on the faint end. By normalizing on the bright galaxies (${\rm 12 < r < 19}$) detected in the central frame (this has a radius of about ${\rm 0.27 h^{-1}}$ Mpc) we derive a Gaussian luminosity ${\rm L_{G} = 1.73\cdot 10^{11}L_{\odot}}$ and a total luminosity within an Abell Radius ${\rm L_{Tot AR} = 2.16\cdot 10^{11}L_{\odot}}$. Due to the small contribution by the Schechter Luminosity (for all clusters the Luminosity is given in the magnitude range ${\rm M=-24}$ to ${\rm M=-12}$) by using ${\rm \alpha = -1.4}$ rather than ${\rm \alpha = -1.65}$ the luminosity decreases of 7\%. A smaller value, however, is derived using the maximum likelihood fit. We adopt ${\rm L_{Tot AR} = 2.16\cdot 10^{11}L_{\odot}}$.

\begin{figure}[!ht]
\centering
\includegraphics[width=8.5cm]{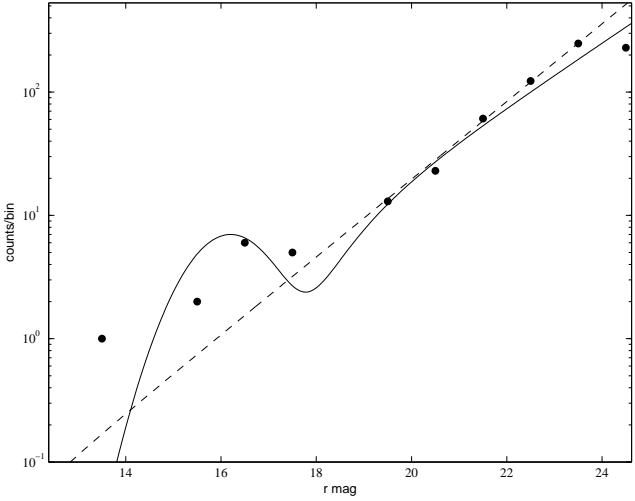}
\caption{A0133. Black dots show that the separation between the Gaussian and the Schechter is not that clear and eventually could also be fitted, with similar uncertainties, with a Schechter.}
\label{fig:11}
\end{figure}

\subsubsection{A4038}

The luminosity of A4038 has been computed using the fit to the bright objects, while for the parameter of the Schechter function component we used the mean, when applicable, of the values estimated for the three colors by using the Maximum Likelihood function, that is: ${\rm \alpha= -1.39}$, ${\rm M^{*}(r) = -13.81+5\log(h)}$. For the ratio between the Gaussian and Schechter part of the LF, we compute ${\rm G/S  = 0.14}$. This value could be strongly dependent on the contamination due to background galaxies, so that we give as a reference also the luminosity computed for ${\rm G/S = 0.8}$ in order to estimate the maximum uncertainties we may have in the result. For the Gaussian fit we use ${\rm Norm = 300}$, ${\rm m_{G} = 16.0}$ and ${\rm \sigma_{G} = 1.0}$.

In the range ${\rm 12 < r < 19}$, that is ${\rm -22.7 < M_{r} <-15.7}$, we observe 30 galaxies in the central field of the cluster. Subtracting the  galaxies in the outskirt of the cluster we have a total of 26, while subtracting the background counts as estimate by \cite{Ty1992} we are left with 20 galaxies. These differences are minors and affect the computation of the cluster luminosity in the central field of only a few percent, so that they could be easily disregarded with respect to other uncertainties. For normalization, we use ${\rm Nobs = 26}$. We also tested for uncertainties due to the estimate of the magnitude of ${\rm M^{*}}$ in the Schechter LF. Field contamination would indeed make this estimate uncertain and favor fainter magnitudes. The total luminosity changes by at most 5\% by brightening ${\rm M^{*}}$ of about 2 magnitudes, these value are given below in parenthesis. We conclude that the estimate of the cluster luminosity, dominated by the bright galaxies, is rather robust. The estimate gives: ${\rm L_{cluster} = 6.7\cdot 10^{10}}$ (${\rm 7.8\cdot 10^{10}}$) ${\rm L_{\odot}}$ with a contribution by the bright galaxies of 98\% (84\%). If we change the ratio the two Luminosity Functions, that is making ${\rm G/S =0.5}$ for instance, the contribution of the Schechter part is reduced to about 5\% while the contribution of the bright galaxies remains constant.

\begin{figure}[!ht]
\centering
\includegraphics[width=8.5cm]{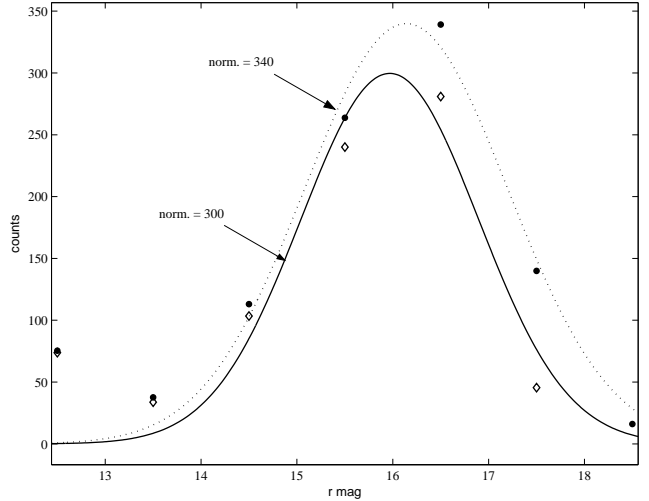}
\caption{A4038. The fit using the counts (black dots) obtained subtracting the outskirt cluster region from the central region gives a normalization value of 350 counts per magnitude per square degree. The counts obtained by subtracting the background as estimated by \cite{Ty1992}, diamonds, give a somewhat smaller value. We adopt a normalization factor equal to 300.}
\label{fig:12}
\end{figure}

For this cluster we normalized on the number of objects that have a distance from the cluster center that is smaller than 400$"$. Using the adopted parameters, we estimate that the luminosity due to faint galaxies within an Hubble radius is ${\rm L_{faint} = 3.7\cdot10^{9} L_{\odot}}$. Since there is no additional contribution by bright galaxies, the toatal luminosity of the cluster would be ${\rm L_{Total} = 7.0\cdot10^{10} L_{\odot}}$. Assuming the knee of the Schechter LF to be 2 magnitudes brighter, we derive a somewhat brighter luminosity, ${\rm L_{Total} = 9.8\cdot 10^{10} L_{\odot}}$. In this case, however, the number of expected galaxies would have been larger (525 rather than 139) and we would have detected this larger over density. We adopt ${\rm L_{TotAR} = 7.0\cdot 10^{10} L_{\odot}}$.\\
The cluster shows three galaxies brighter than apparent magnitude 14.0 which do not differ much in luminosity. The brightest galaxy has a luminosity ${\rm L_{g} =  0.67\cdot10^{11}L_{\odot}}$, ${\rm L_{r} = 0.51\cdot 10^{11}L_{\odot}}$, ${\rm L_{i} = 0.39\cdot 10^{11}L_{\odot}}$. By adding the three brightest galaxies we have: ${\rm L_{g} = 2.12\cdot 10^{11}L_{\odot}}$, ${\rm L_{r} = 1.68\cdot 10^{11}L_{\odot}}$, ${\rm L_{i} = 1.56\cdot 10^{11}L_{\odot}}$. And this clearly shows: 
\begin{itemize}
\item[a)] that most of the light is contained in the central bright galaxies; 
\item[b)] that sometimes we can not measure the cluster luminosity simply by fitting the counts.
\end{itemize}

\subsubsection{A4059}

As discussed in Sec. \ref{sec:4059}, the different estimates of the parameters of the LF for A4059 are not in perfect agreement and depend somewhat of the method of analysis. We adopted those, as we illustrated there, which better satisfy the visual inspection and, in spite of smaller statistics, have been derived after careful sky subtraction. Also in this case the cD does not to fit properly the luminosity distribution function: its luminosity indeed is larger than the whole Gaussian contribution so that we should correct for this later on. The luminosity of the cD is $L_{g} =  1.72\cdot 10^{11} L_{\odot}$, $L_{r} =  1.88\cdot10^{11}L_{\odot}$, $L_{i} = 1.56\cdot10^{11}L_{\odot}$. The total cluster luminosity is $L_{TotAR} \sim 1.0\cdot 10^{11} L/L_{\odot}$ and $L_{TotAR} = 1.9\cdot 10^{11} L/L_{\odot}$ applying a correction for the cD.

The luminosity of the core without accounting for the cD galaxy and using both the adopted distribution function and the others as derived in the respective thesis work are: ${\rm L_{core} = 7.5\cdot 10^{10}L_{\odot}}$; ${\rm L_{core}^{Ratti K.} = 4.9\cdot 10^{10}L_{\odot}}$; ${\rm L_{core}^{Parolin I.} = 4.4\cdot 10^{10}L_{\odot}}$.

\subsection{The luminosity within 400$''$ from the cluster center}

To check for consistency with the total luminosity computed using the parameters of the LF and the King's profile, we also measured the luminosity of the central field of each cluster, that is within 400$''$ from the cluster center of the observed area, due to galaxies brighter than ${\rm M_{g} = -16.5}$, ${\rm M_{r} = -17.0}$, and ${\rm M_{i} = -17.0}$ by simply adding up the luminosity of the single galaxies. To the luminosity so computed, we subtracted the contribution of the background, Eq. \ref{eq:bkg}, with:
\begin{equation}
\begin{array}{lcccr}
{\rm N_{g}=5128} & , & {\rm N_{r}=6680} & , & {\rm N_{i}=7920}\\
\end{array}
\end{equation}

and the luminosity of the background computed as:

\begin{equation}
{\rm \int^{m(M_{g}=-16.5, M_{r,i}= -17)}_{m=11.5}N(m)L(m)dm}
\end{equation}

since we did not detect cluster galaxies brighter that $m=11.5$ in any filter.

 Only for the cluster EXO 0422-086 we used magnitude corrected for galactic extinction. The cluster magnitudes estimated in this way are not affected by any bias due to the analysis. However, more distant clusters (as for instance A3695) will appear brighter since a larger cluster area has been accounted for. The effect is however rather small, and we can account for it anyway, for we are considering only fairly bright galaxies. Trying to have results as robust as possible, we computed also the Cluster Luminosity as follows:
using the estimate of the LF derived by the Maximum Likelihood method we computed the luminosity of the central region observed, Field 1, normalizing to the number of galaxies observed in this Field. The Field covers the observed galaxies within ${\rm R < 400}$ arcs and, except for the cluster A3695, is of the order of the core radius. We add the contribution of the brightest galaxies excluded from the fit: only one with ${\rm m < 14.0}$ for all the clusters but A3695, for which the fit was carried out for galaxies $m > 15.5$. To compute the cluster luminosity to the Abell radius ($1.5 h^{-1}$ Mpc), as we indicated before, we add the contribution of the Schechter function component weighted by a King's profile for accounting of the density distribution. We used a core radius of about 0.25 Mpc except in the case of A0085 and A4059, clusters for which we were able to measure a core radius, albeit with rather large errors. For each cluster, we subtracted from the galaxy counts the background interpolating for the filters we used the counts given by \cite{Ty1992} in different colors.

As it was mentioned before, the Maximun likelihood fit uses the galaxy sequence and assumes that here the cluster members dominate, because we use the sequence galaxies within the central 400$''$. In other words, we do not subtract the background and have enough statistics for a reasonable analysis.

\section{Summary and conclusions}

For the sample of clusters we discussed in this paper, the optical luminosity we estimated, Tab. \ref{tab:lums}, correlates reasonably well with the luminosity observed in the X-ray band (data from Strubel \& Rood, 1999). This is a well known result and we also observe that there is a tendency for the cluster luminosity to correlate with the redshift. Indeed by going at large distances any sample tend to pick up brighter objects. The cluster X-ray luminosity of the sample correlates as well with the redshift and that is also expected for the reason we just said.

\begin{figure}[!ht]
\centering
\includegraphics[width=8.5cm]{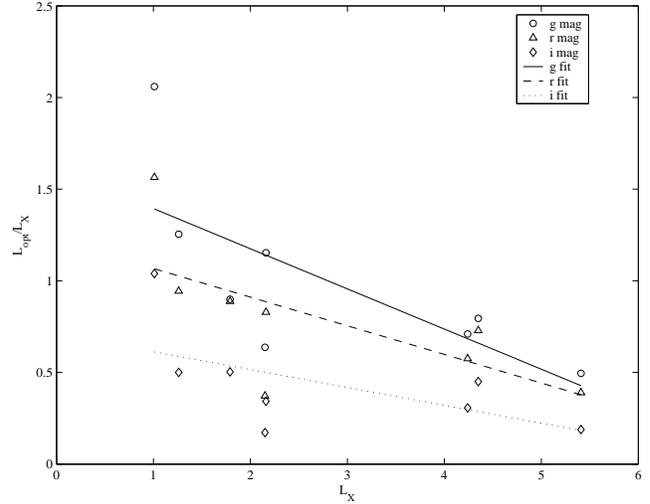}
\caption{The ratio of the optical Luminosity (in the \textit{g}, \textit{r} and \textit{i} bands) of the clusters to their X-ray luminosity versus the X-ray luminosity. Brighter X-ray clusters are, percent wise, less luminous than fainter X-ray clusters. Furthermore the slope of the correlations seems to indicate that the cluster colors are a function of the X-ray luminosity in the sense that bright X-ray clusters are redder.}
\label{fig:13}
\end{figure}

The important result we find, albeit with rather small statistical significance due to the small span in luminosityes, is that the most luminous clusters in the X-ray band are comparatively optically fainter than those having smaller luminosity in the X-ray band. The effect is rather robust if we consider the fact that it is manifest in all three fitlers. Indeed in Fig. \ref{fig:13} we plot the ratio ${\rm L_{opt}/L_{X}}$ versus ${\rm L_{X}}$ and the effect is evident: the ratio ${\rm L_{opt}/L_{X}}$ decreases as a function of ${\rm L_{X}}$. The slope of the correlation, furthermore, decerases as a function of the wavelength, that is bright X-ray clusters are redder than fainter X-ray clusters.

The question arises whether the effect we observe is due to a transfer of matter between the galaxies and the intracluster medium (ICM) where the bright X-ray clusters had an higher stellar activity in the past. Some of these points will be discussed in a subsequent paper.

In summay, and over the years, we measured a set of clusters in various colors with the purpose of better understanding the shape of the luminosity function in clusters and to pin point the slope of the faint end. The observations were carried out in different colors and the analysis, and the interpretation of the data, made use of different algorithms to make sure that we were not biased by the method of analysis. All this work, while respectable, lead to a rather robust bi-modal cluster luminosity function, a result that is not new. Given the sample we used we were unable to estimate any variation of the LF as a function of the distance of the cluster center. While we find indication in the literature that there are variations (here we refer to the slope of the faint end), we feel that such results, since it is very important for the understanding of the cluster dynamics and cluster evolution, should be checked on large spectroscopic samples.

The important result of this work, however, consists in the estimate of the total luminosity in different optical color and its correlation with the X-ray luminosity. Indeed we believe that it is the estimate of the luminosity and colors of clusters that will give the opportunity, when compared to the X-ray luminosity, to better pin point some of the characteristics of the cluster and ICM evolution.

\begin{acknowledgements}
We thank Alberto Moretti for the many discussions in the course of this work and the referee,  P. Schuecker, for the  helpful suggestions.
\end{acknowledgements}

\onecolumn

\begin{table*}[!h]
\centering
\begin{tabular}{|c|c|c|c|c|c|c|}\hline
Cluster&Filter&G/S&${\rm m_{G}}$&${\rm \sigma_{G}}$&${\rm m^{*}}$&${\rm \alpha}$\\\hline\hline
&&&&&&\\
&g&$3.50^{+3.50}_{-1.50}$&$16.01^{+0.24}_{-0.24}$&$0.29^{+0.16}_{-0.16}$&$15.30^{+0.90}_{-1.80}$&$-1.59^{+0.04}_{-0.04}$\\
&&&&&&\\\cline{2-7}
&&&&&&\\
A0085&r&$3.70^{+4.30}_{-2.10}$&$15.63^{+0.27}_{-0.27}$&$0.34^{+0.14}_{-0.14}$&$15.05^{+0.97}_{-2.80}$&$-1.59^{+0.04}_{-0.04}$\\
&&&&&&\\\cline{2-7}
&&&&&&\\
&i&$3.67^{+4.30}_{-2.03}$&$15.37^{+0.20}_{-0.20}$&$0.33^{+0.16}_{-0.16}$&$14.44^{+0.60}_{-3.20}$&$-1.55^{+0.04}_{-0.04}$\\
&&&&&&\\\hline
&&&&&&\\
&g&$0.81^{+0.27}_{-0.27}$&$17.07^{+0.18}_{-0.18}$&$0.82^{+0.19}_{-0.19}$&$19.25^{+0.36}_{-0.36}$&$-1.71^{+0.07}_{-0.07}$\\
&&&&&&\\\cline{2-7}
&&&&&&\\
A0133&r&$0.29^{+0.8}_{-0.1}$&$16.84^{+0.31}_{-0.31}$&$0.85^{+0.5}_{-0.2}$&$19.69^{+0.9}_{-0.2}$&$-1.34^{+0.17}_{-0.17}$\\
&&&&&&\\\cline{2-7}
&&&&&&\\
&i&$0.34^{+0.16}_{-0.16}$&$16.69^{+0.26}_{-0.26}$&$0.88^{+0.27}_{-0.27}$&$19.28^{+0.5}_{-0.2}$&$-1.59^{+0.07}_{-0.07}$\\
&&&&&&\\\hline
&&&&&&\\
&g&$0.34^{+0.1}_{-0.1}$&$17.22^{+0.24}_{-0.24}$&$1.04^{+0.23}_{-0.23}$&$19.23^{+0.30}_{-0.30}$&$-1.42^{+0.08}_{-0.08}$\\
&&&&&&\\\cline{2-7}
&&&&&&\\
A3667&r&$0.16^{+0.04}_{-0.04}$&$16.66^{+0.31}_{-0.31}$&$0.97^{+0.27}_{-0.27}$&$18.67^{+0.48}_{-0.48}$&$-1.39^{+0.09}_{-0.09}$\\
&&&&&&\\\cline{2-7}
&&&&&&\\
&i&$0.5^{+0.15}_{-0.15}$&$16.32^{+0.48}_{-0.48}$&$1.03^{+0.59}_{-0.18}$&$18.22^{+0.56}_{-1.0}$&$-1.52^{+0.07}_{-0.07}$\\
&&&&&&\\\hline
&&&&&&\\
&g&$0.8^{+0.27}_{-0.5}$&$18.52^{+0.3}_{-0.3}$&$1.03^{+0.23}_{-0.23}$&$19.69^{+0.89}_{-0.89}$&$-1.63^{+0.14}_{-0.14}$\\
&&&&&&\\\cline{2-7}
&&&&&&\\
A3695&r&$0.95^{+0.86}_{-0.12}$&$17.90^{+0.28}_{-0.28}$&$0.93^{+0.20}_{-0.20}$&$19.16^{+0.62}_{-0.62}$&$-1.68^{+0.10}_{-0.10}$\\ 
&&&&&&\\\cline{2-7}
&&&&&&\\
&i&$0.38^{+0.09}_{-0.09}$&$17.88^{+0.32}_{-0.32}$&$0.90^{+0.22}_{-0.22}$&$19.93^{+0.33}_{-0.33}$&$-1.54^{+0.06}_{-0.06}$\\
&&&&&&\\\hline
\end{tabular}
\caption{Parameters obtained with Maximum Likelihood method and the standard analysis}
\label{tab:riski1}
\end{table*}
\onecolumn
\begin{table*}[!h]
\centering
\begin{tabular}{|c|c|c|c|c|c|c|c|}\hline
Cluster&Filter&G/S&${\rm m_{G}}$&${\rm \sigma_{G}}$&${\rm m^{*}}$&${\rm \alpha}$\\\hline\hline
&&&&&&\\
&g&$0.14^{+0.05}_{-0.05}$&$17.14^{+0.40}_{-0.40}$&$1.72^{+0.31}_{-0.31}$&$21.12^{+0.20}_{-0.20}$&$-1.39^{+0.07}_{-0.07}$\\
&&&&&&\\\cline{2-7}
&&&&&&\\
A4038&r&$0.14^{+0.05}_{-0.05}$&$16.54^{+0.58}_{-0.58}$&$2.02^{+0.59}_{-0.59}$&$20.88^{+0.23}_{-0.23}$&$-1.38^{+0.07}_{-0.07}$\\
&&&&&&\\\cline{2-7}
&&&&&&\\
&i&$0.14^{+0.04}_{-0.04}$&$16.47^{+0.53}_{-0.53}$&$1.86^{+0.49}_{-0.49}$&$20.86^{+0.19}_{-0.19}$&$-1.41^{+0.06}_{-0.06}$\\
&&&&&&\\\hline
&&&&&&\\
&g&$1.16^{+0.70}_{-0.70}$&$18.20^{+0.84}_{-0.84}$&$1.34^{+0.80}_{-0.40}$&$19.06^{+1.30}_{-1.50}$&$-1.74^{+0.12}_{-0.12}$\\
&&&&&&\\\cline{2-7}
&&&&&&\\
A4059&r&$0.76^{+4.20}_{-0.30}$&$17.22^{+0.95}_{-0.95}$&$1.35^{+1.40}_{-0.60}$&$18.07^{+1.00}_{-3.10}$&$-1.65^{+0.70}_{-0.10}$\\
&&&&&&\\\cline{2-7}
&&&&&&\\
&i&$1.79^{+4.20}_{-0.30}$&$16.54^{+3.00}_{-0.40}$&$1.08^{+4.20}_{-0.30}$&$17.87^{+2.20}_{-3.50}$&$-1.67^{+0.11}_{-0.11}$\\
&&&&&&\\\hline
&&&&&&\\
&g&$0.55^{+0.27}_{-0.16}$&$17.82^{+4.2}_{-0.76}$&$1.64^{+2.13}_{-.65}$&$19.26^{+1.53}_{-0.42}$&$-1.66^{+0.06}_{-0.06}$\\
&&&&&&\\\cline{2-7}
&&&&&&\\
EXO 0422-086&r&$0.43^{+0.20}_{-0.20}$&$17.46^{+0.24}_{-0.24}$&$0.86^{+0.20}_{-0.20}$&$18.98^{+0.36}_{-0.36}$&$-1.47^{+0.06}_{-0.06}$\\
&&&&&&\\\cline{2-7}
&&&&&&\\
&i&$0.38^{+0.13}_{-0.13}$&$17.26^{+0.28}_{-0.28}$&$0.89^{+0.20}_{-0.20}$&$18.39^{+0.53}_{-0.53}$&$-1.40^{+0.07}_{-0.07}$\\
&&&&&&\\\hline
&&&&&&\\
&g&$0.21^{+0.85}_{-0.06}$&$17.21^{+0.28}_{-0.28}$&$0.91^{+0.32}_{-0.32}$&$18.97^{+0.9}_{-2.0}$&$-1.34^{+0.17}_{-0.17}$\\
&&&&&&\\\cline{2-7}
&&&&&&\\
A0496&r&$0.22^{+0.06}_{-0.06}$&$16.80^{+0.33}_{-0.33}$&$0.98^{+0.31}_{-0.31}$&$18.31^{+0.6}_{-0.6}$&$-1.68^{+0.08}_{-0.08}$\\
&&&&&&\\\cline{2-7}
&&&&&&\\
&i&$0.78^{+0.71}_{-0.36}$&$16.46^{+0.21}_{-0.21}$&$0.86^{+0.20}_{-0.20}$&$18.38^{+0.6}_{-1.1}$&$-1.43^{+0.16}_{-0.93}$\\
&&&&&&\\\hline
\end{tabular}
\caption{Parameters obtained with Maximum Likelihood method and the standard analysis}
\label{tab:riski2}
\end{table*}

\onecolumn

\begin{table*}[!h]
\centering
\begin{tabular}{|c|c|c|c|c|c|c|c|c|c|}\hline
Cluster&z&${\rm L_{X}}$&Filter&${\rm L_{0}/L_{\odot}}$&${\rm L_{0}/L_{\odot}}$&${\rm L_{0}/L_{\odot}}$&${\rm L_{0}/L_{\odot}}$&${\rm L/L_{X}}$&${\rm L/L_{X}}$\\
&&${\rm 10^{44} h^{-2}}$erg/s&&${\rm 10^{11}h^{-2}}$erg/s&${\rm 10^{11}h^{-2}}$erg/s&${\rm 10^{11}h^{-2}}$erg/s&${\rm 10^{11}h^{-2}}$erg/s&${\rm F1_{obs}}$&A.R.\\
&&&&${\rm F1_{obs}}$&${\rm F1_{com}}$&0.3Mpc&A.R.&&\\\hline\hline
&&&g&2.68&2.81&2.94&3.68&0.4954&0.6802\\ \cline{4-10}
A0085&0.0556&5.41&r&2.11&2.59&2.69&3.22&0.3900&0.5952\\ \cline{4-10} 
&&&i&1.02&1.69&1.94&1.94&0.1885&0.3586\\\hline
&&&g&2.49&2.84&2.86&3.01&1.1528&1.3935\\ \cline{4-10}
A0133&0.0566&2.16&r&1.79&1.92&1.93&2.03&0.8287&0.9398\\ \cline{4-10}
&&&i&0.74&1.13&1.14&1.18&0.326&0.5463\\\hline
&&&g&3.46&3.67&3.72&3.89&0.7954&0.8943\\ \cline{4-10}
A3667&0.0556&4.35&r&3.17&3.25&3.32&3.58&0.7287&0.8230\\ \cline{4-10}
&&&i&1.96&2.26&2.27&2.36&0.4506&0.5425\\\hline
&&&g&3.01&5.63&5.54&5.97&0.7099&1.4080\\ \cline{4-10}
A3695&0.0893&4.24&r&2.44&6.06&5.97&6.35&0.5755&1.4976\\ \cline{4-10}
&&&i&1.30&3.86&3.8&4.04&0.3066&0.9528\\\hline
&&&g&2.08&2.83&2.84&2.86&2.0594&2.8317\\ \cline{4-10}
A4038&0.0292&1.01&r&1.58&2.93&2.94&2.95&1.5644&2.9208\\ \cline{4-10}
&&&i&1.05&1.09&1.09&1.09&1.0396&1.0792\\\hline
&&&g&1.61&1.64&1.66&1.72&0.8994&0.9609\\ \cline{4-10}
A4059&0.0460&1.79&r&1.59&1.75&1.76&1.80&0.8883&1.0056\\ \cline{4-10}
&&&i&0.90&1.03&1.03&1.03&0.5028&0.5754\\\hline
&&&g&1.58&1.83&1.86&1.91&1.2540&1.5159\\ \cline{4-10}
EXO 0422-086&0.0397&1.26&r&1.19&0.91&0.92&0.96&0.9444&0.7619\\ \cline{4-10}
&&&i&0.63&0.85&0.87&0.92&0.5000&0.7302\\\hline
&&&g&1.37&1.28&1.37&1.51&0.6372&0.7023\\ \cline{4-10}
A0496&0.0328&2.15&r&0.80&0.99&1.1&1.25&0.3721&0.5814\\ \cline{4-10}
&&&i&0.37&0.60&0.61&0.63&0.1721&0.2930\\\hline
\end{tabular}
\caption{Luminosities obtained with the parameters of the standard analysis. X-ray luminosity computed in the ${\rm 0.5-2.0}$ keV band assuming a power-law spectrum with energy index ${\rm \gamma=0.4}$ (data from Strubel \& Rood 1999). NOTE: the \textit{i} luminosity of A0133, A3695, A4059 and EXO 0422-086 after background correction resulted to be slightly smaller than the luminosity of the few brightest galaxies, due to a sky background overcorrection and uncertainties in the cluster counts. In these cases, the luminosity was corrected assuming that brightest galaxies are cluster members as indicated by the morphology and by the redshift.}
\label{tab:lums}
\end{table*}

\begin{figure}[!ht]
\centering
\begin{tabular}{lr}
\includegraphics[width=9cm, height=11.5cm]{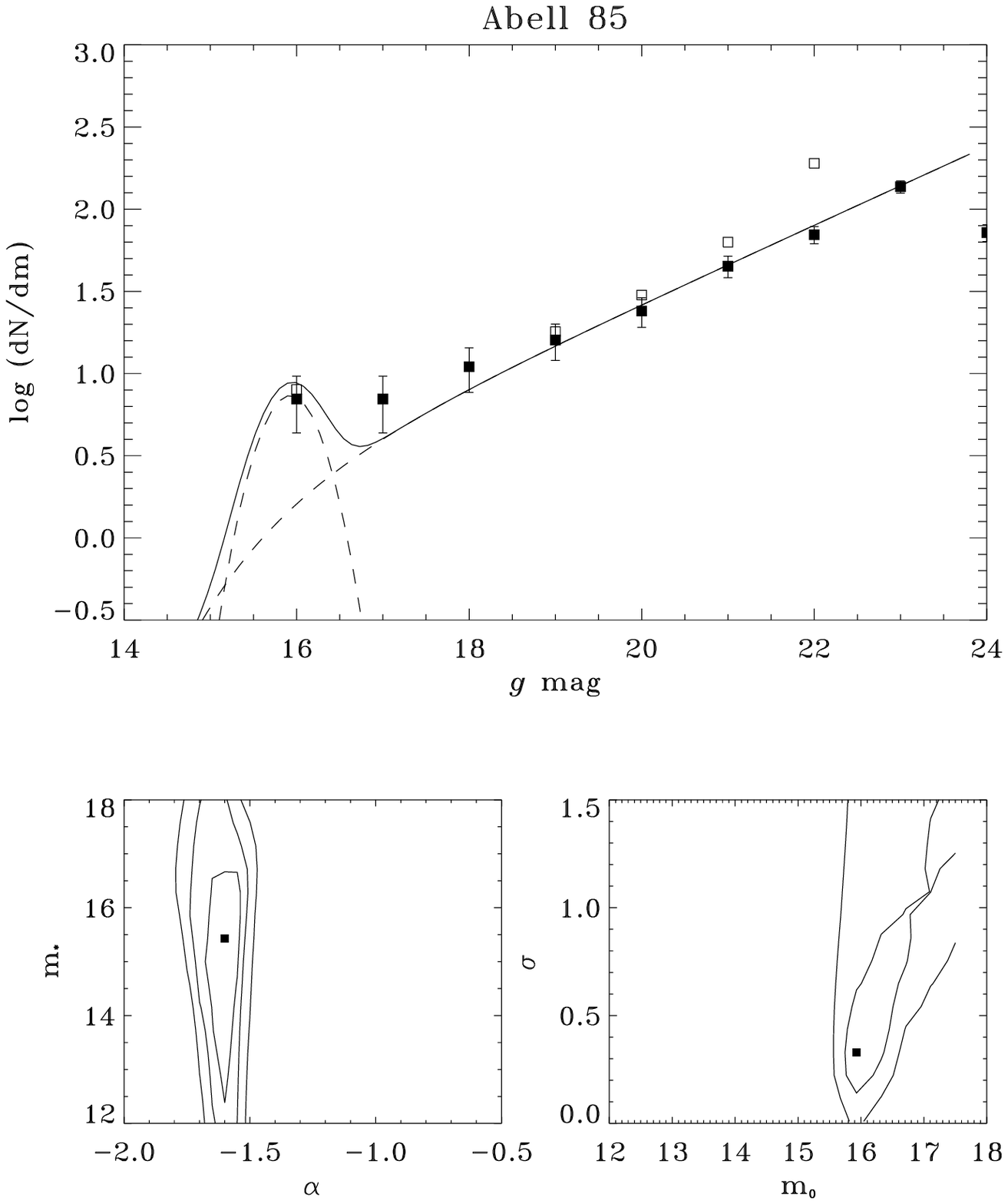} & \includegraphics[width=9cm, height=11.5cm]{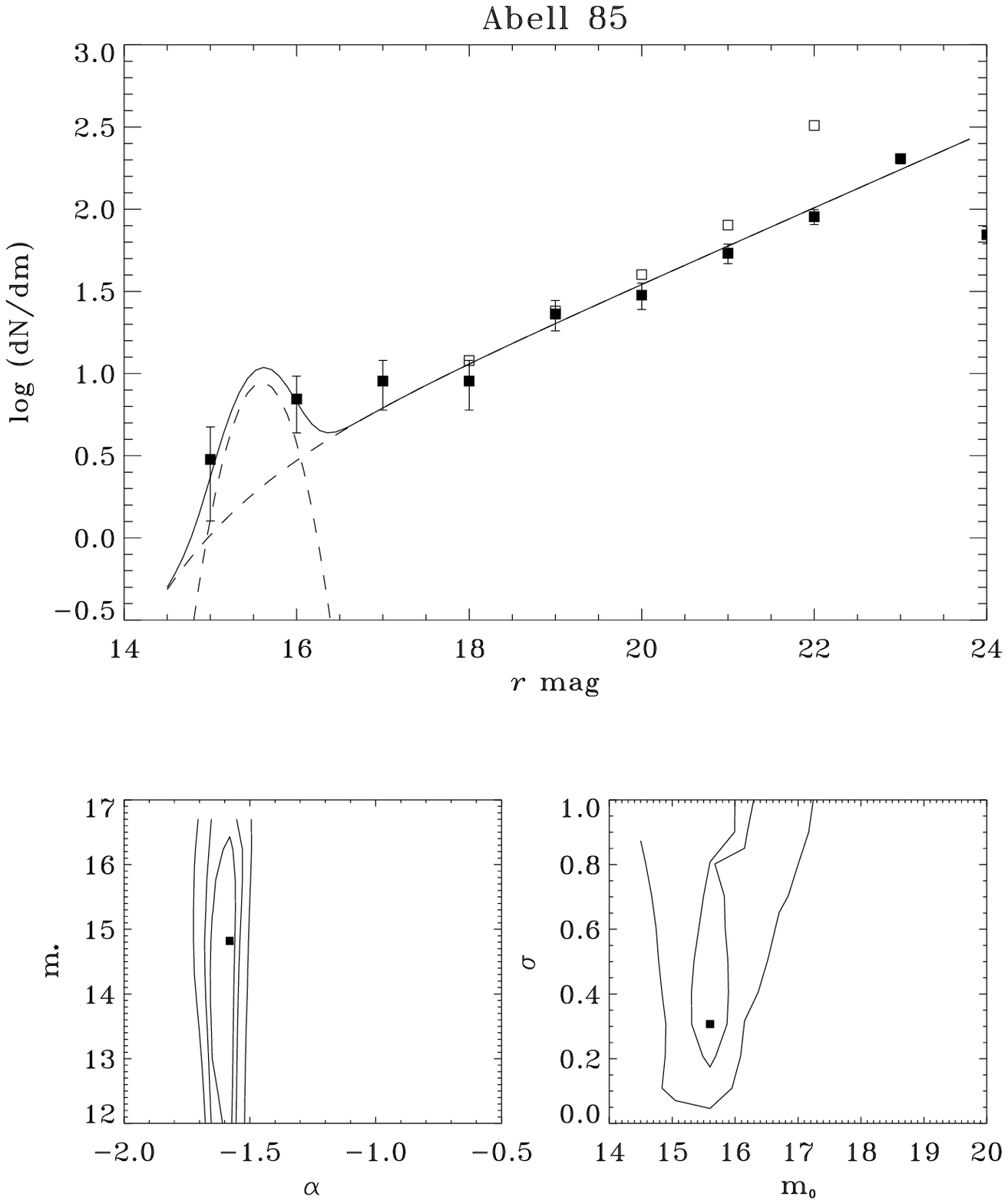} \\
\multicolumn{2}{c}{\includegraphics[width=9cm, height=11.5cm]{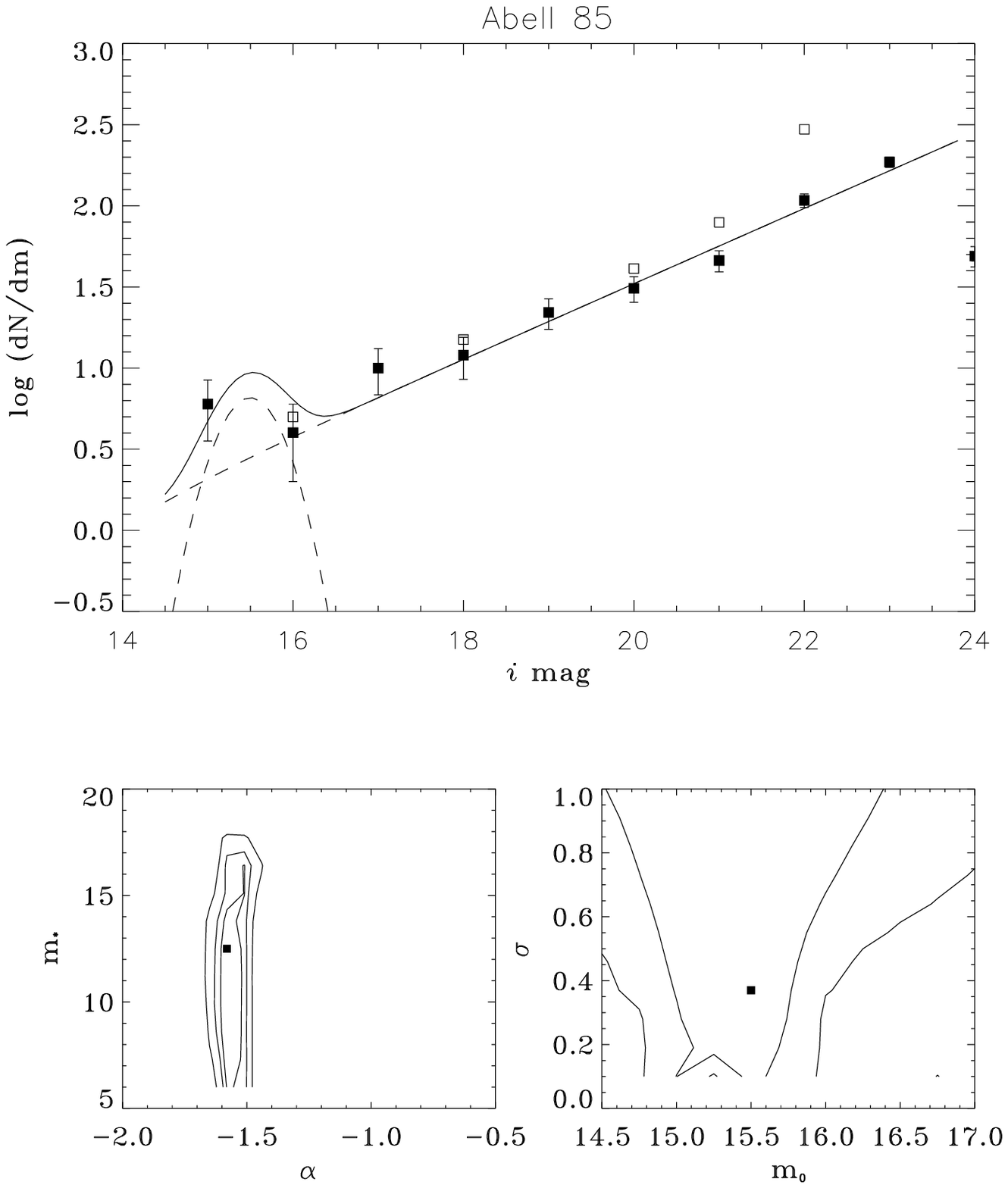}}
\end{tabular}
\caption{Hereafter we show, for each cluster and for each filter used in this work, the fit of our model obtained with the maximum likelihood method and two boxes with the 1${\rm \sigma}$, 2${\rm \sigma}$ and 3${\rm \sigma}$ confidence contours for the parameters of the Schechter ($\alpha$ vs. ${\rm m^{*}}$)  and the Gaussian (${\rm m_{0}}$ vs. ${\rm \sigma}$). In this page: A0085.}
\label{fig:mxl0085}
\end{figure}

\onecolumn

\begin{figure}[!ht]
\centering
\begin{tabular}{lr}
\includegraphics[width=9cm, height=11.5cm]{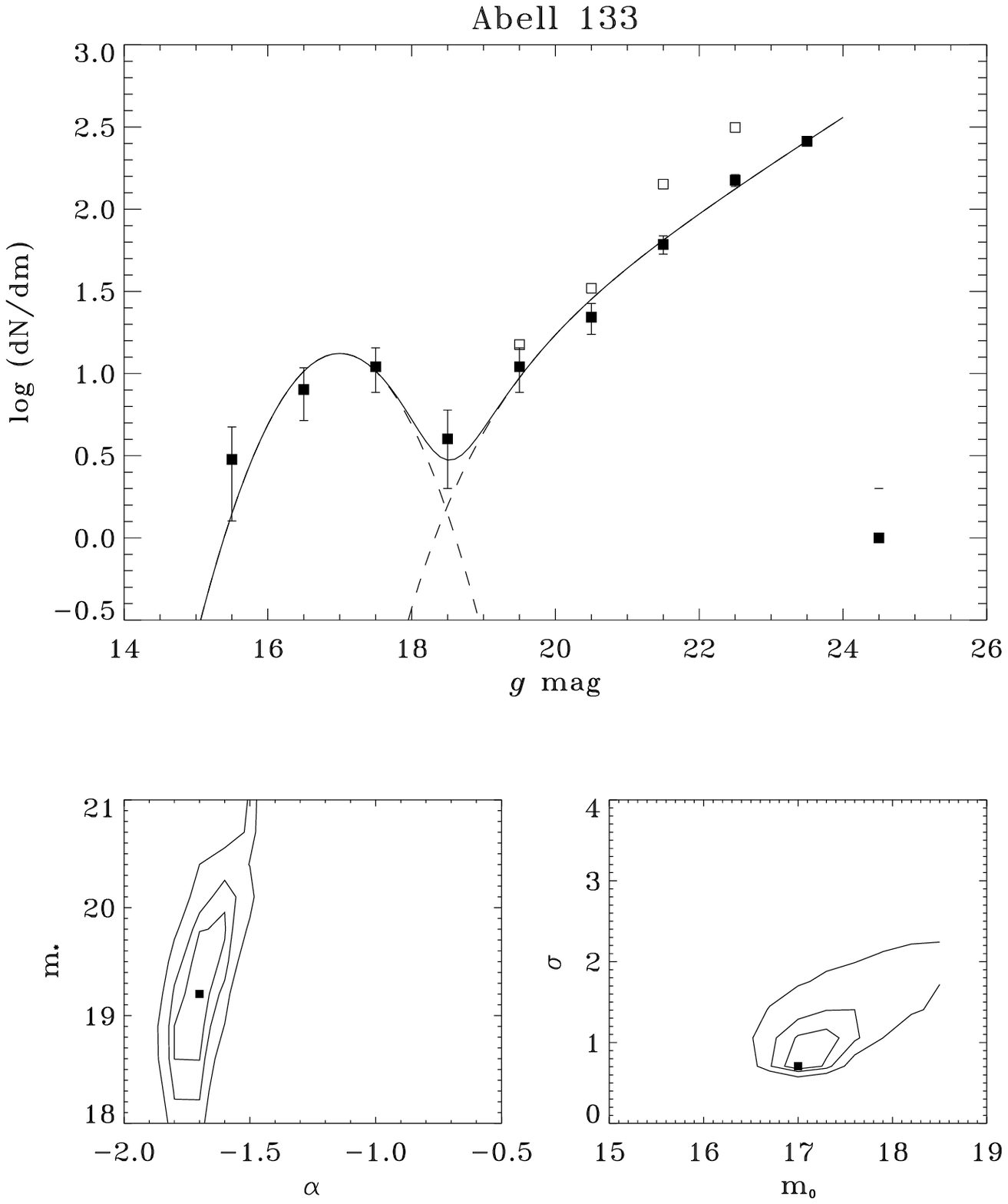} & \includegraphics[width=9cm, height=11.5cm]{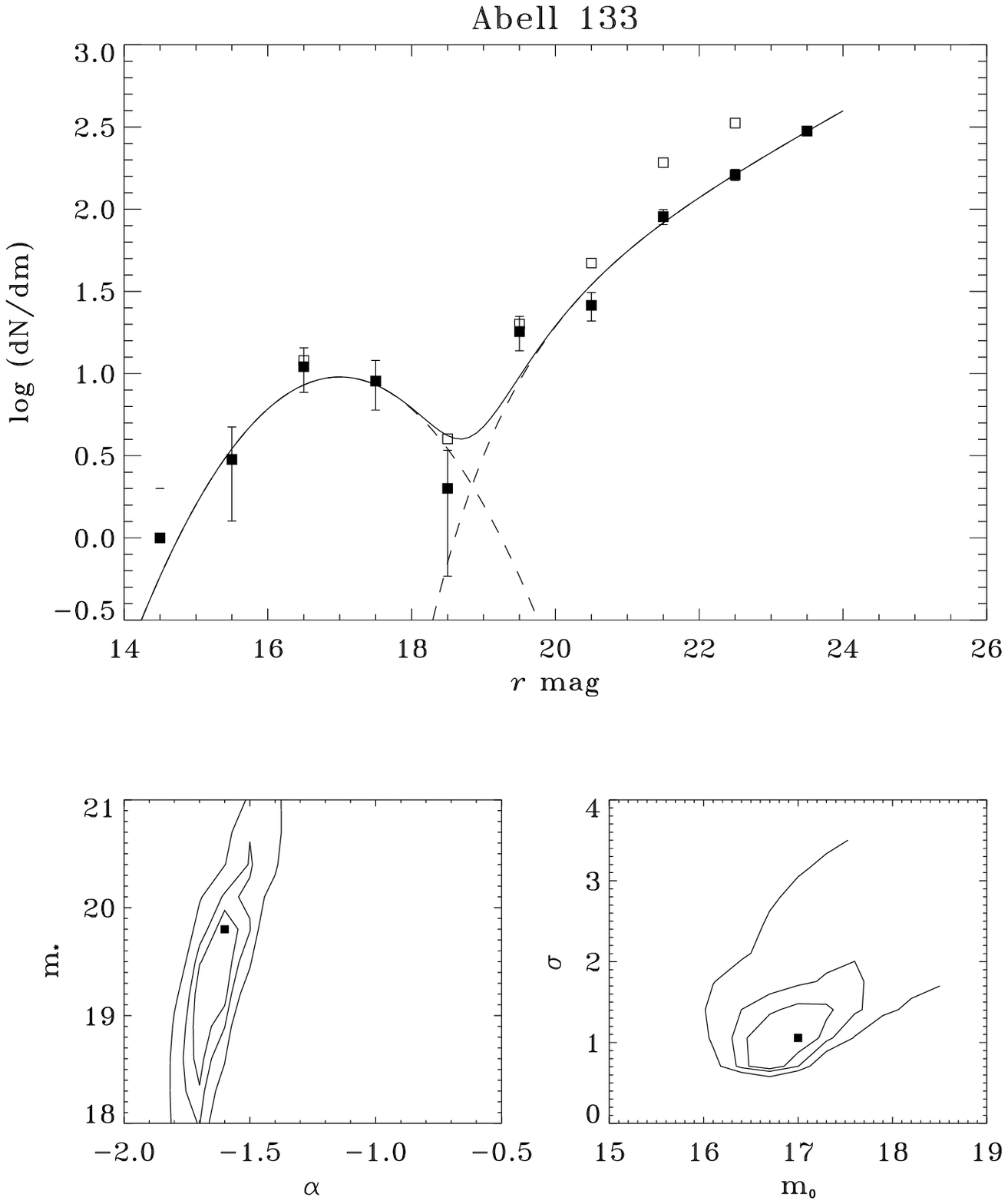} \\
\multicolumn{2}{c}{\includegraphics[width=9cm, height=11.5cm]{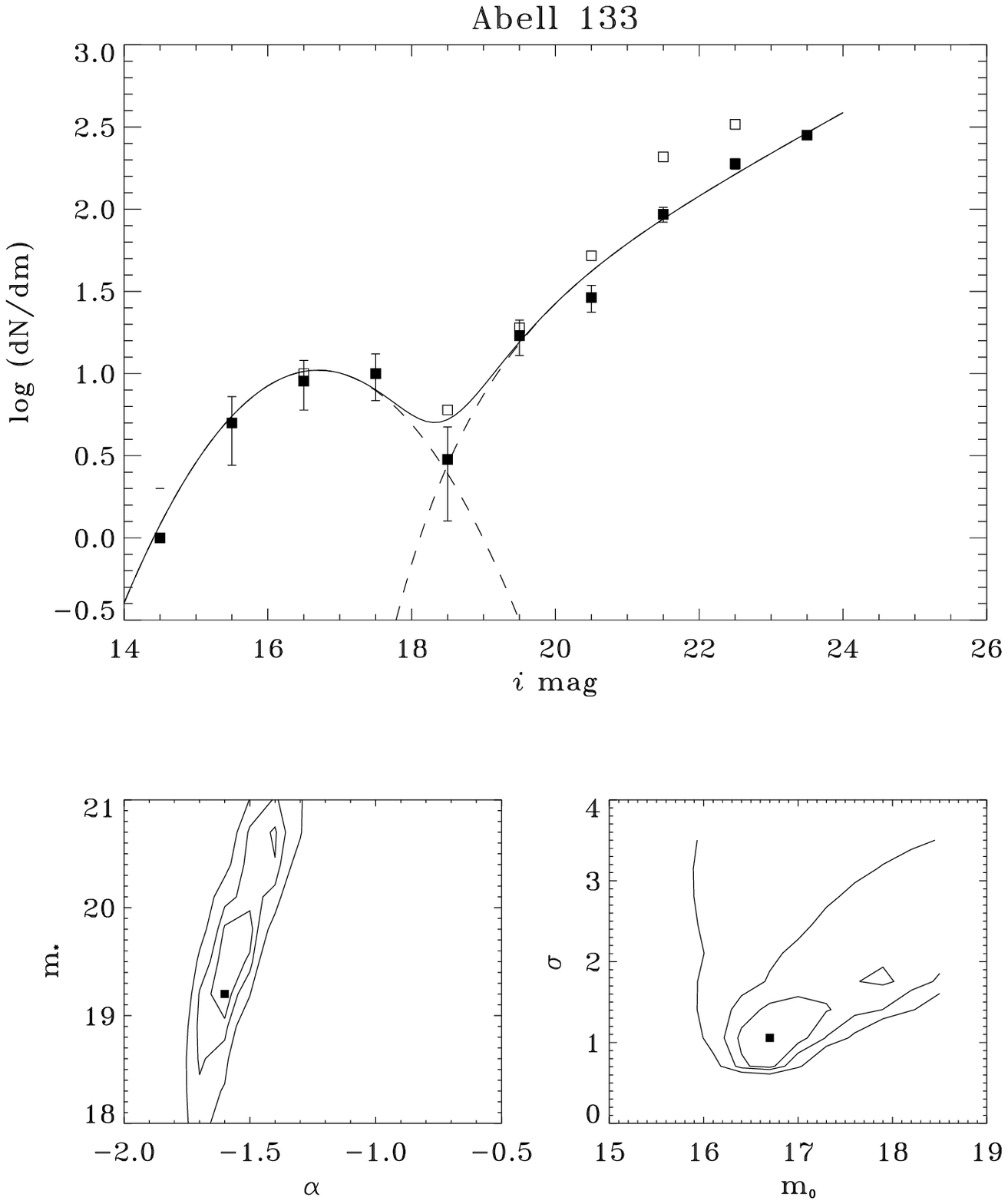}}
\end{tabular}
\caption{A0133}
\label{fig:mxl0133}
\end{figure}

\onecolumn

\begin{figure}[!ht]
\centering
\begin{tabular}{lr}
\includegraphics[width=9cm, height=11.5cm]{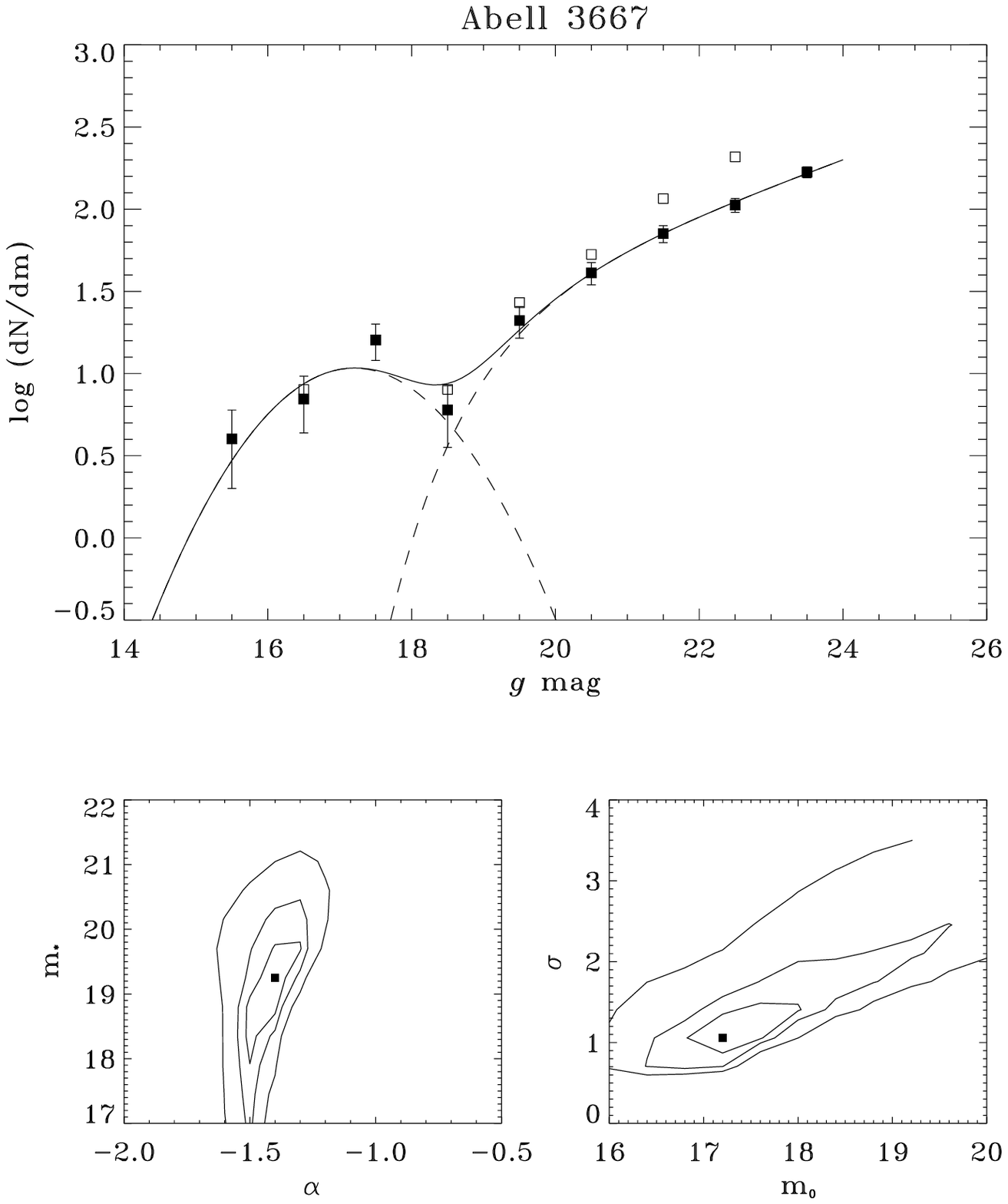} & \includegraphics[width=9cm, height=11.5cm]{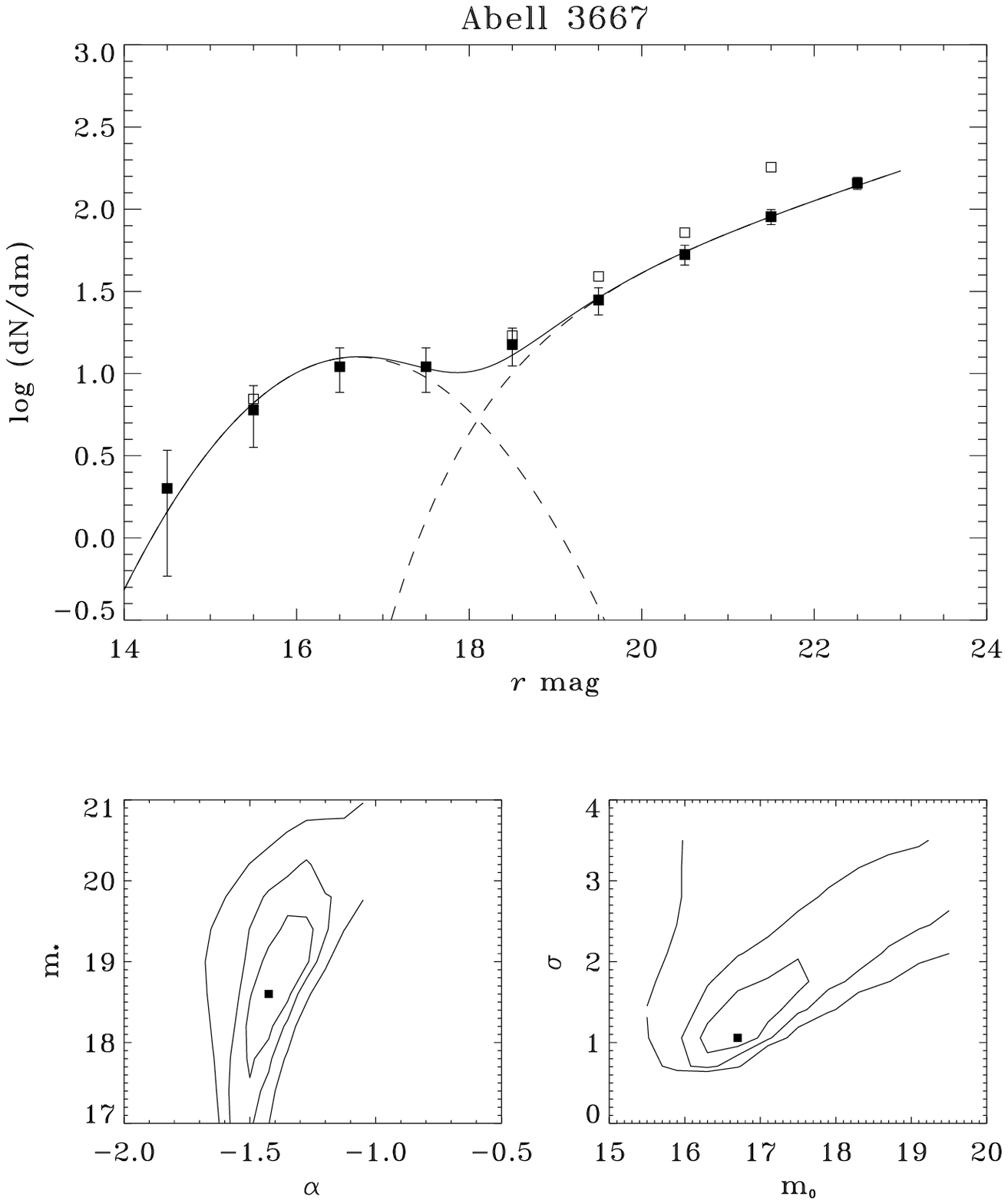} \\
\multicolumn{2}{c}{\includegraphics[width=9cm, height=11.5cm]{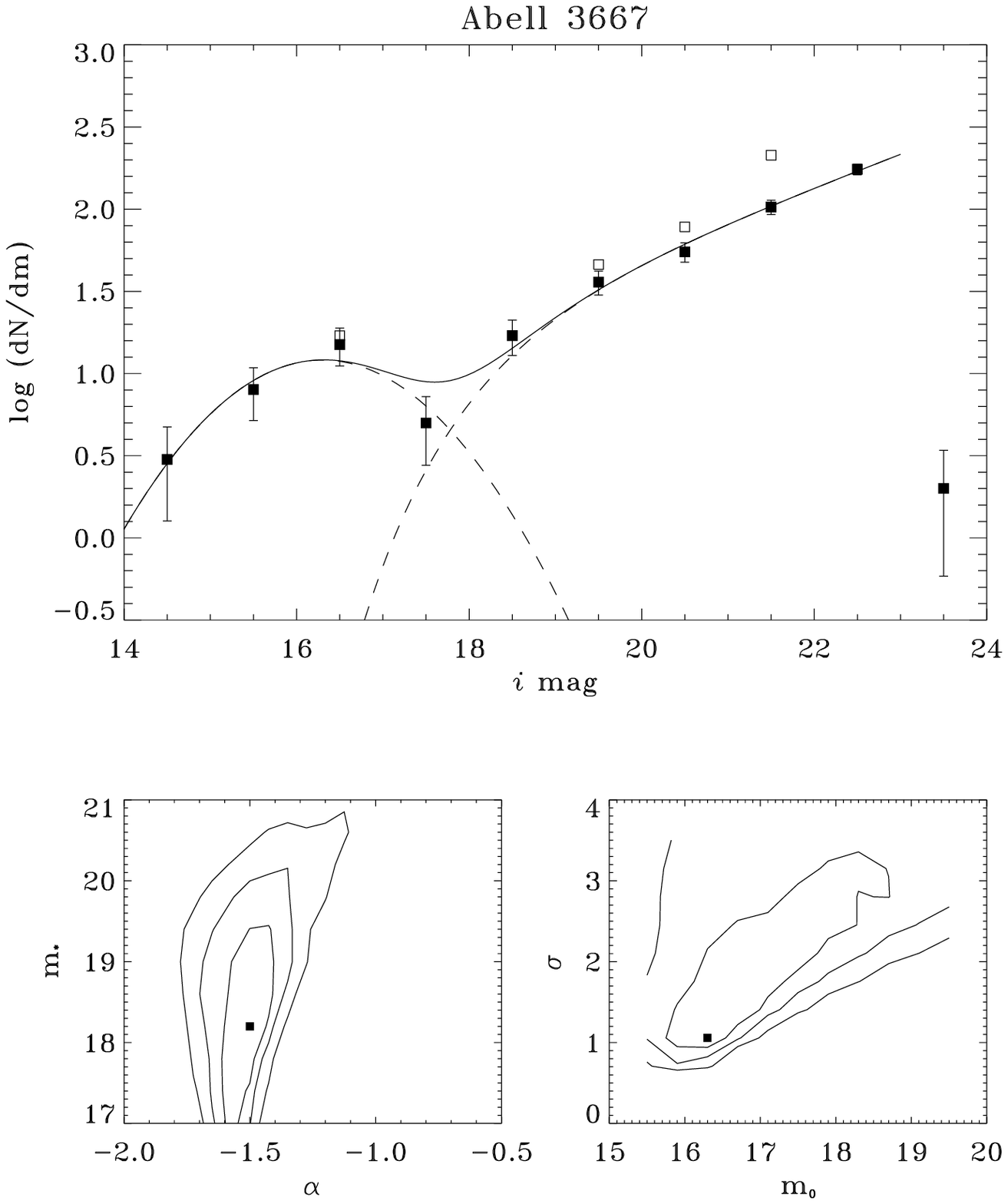}}
\end{tabular}
\caption{A3667}
\label{fig:mxl3667}
\end{figure}

\onecolumn

\begin{figure}[!ht]
\centering
\begin{tabular}{lr}
\includegraphics[width=9cm, height=11.5cm]{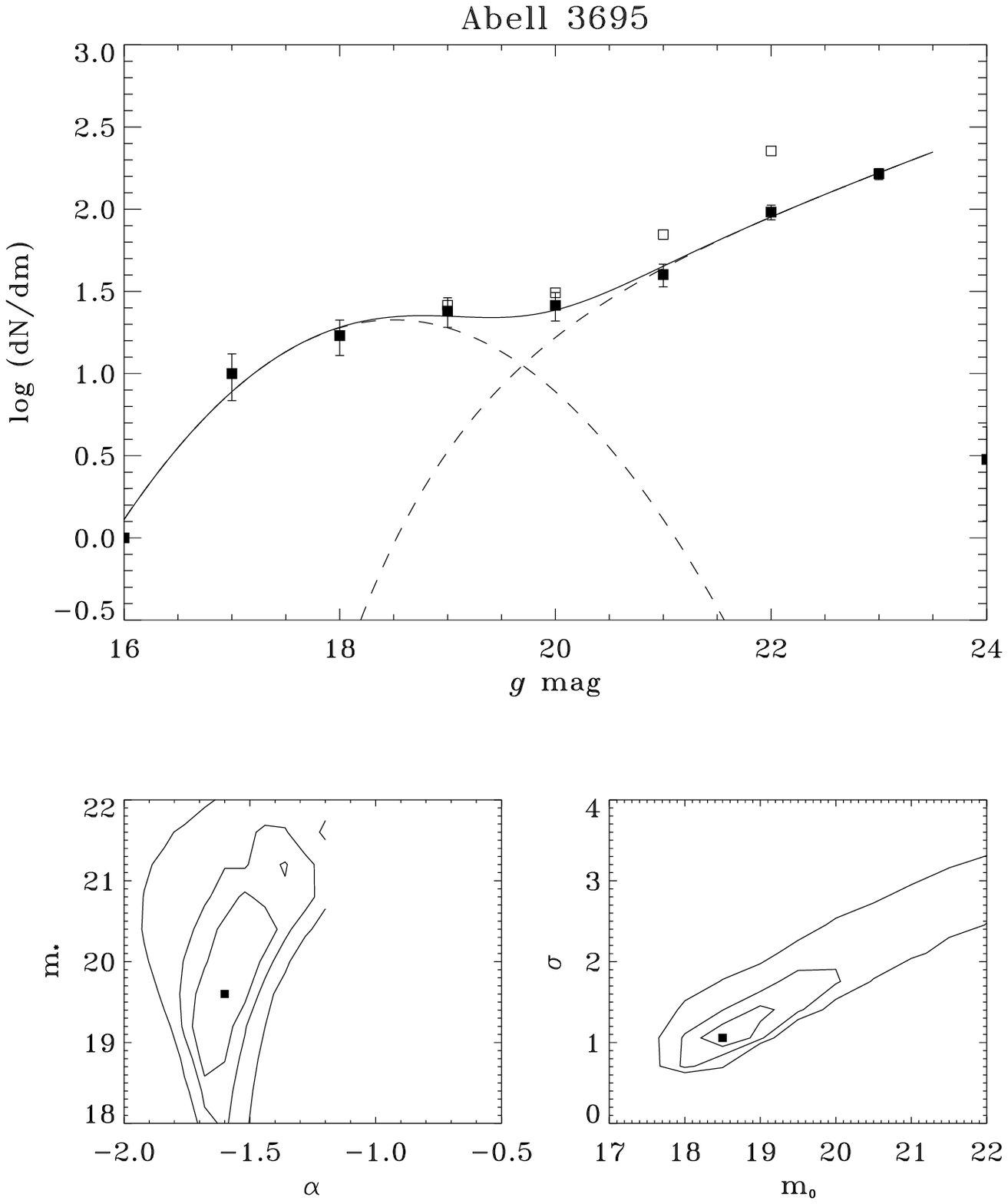} & \includegraphics[width=9cm, height=11.5cm]{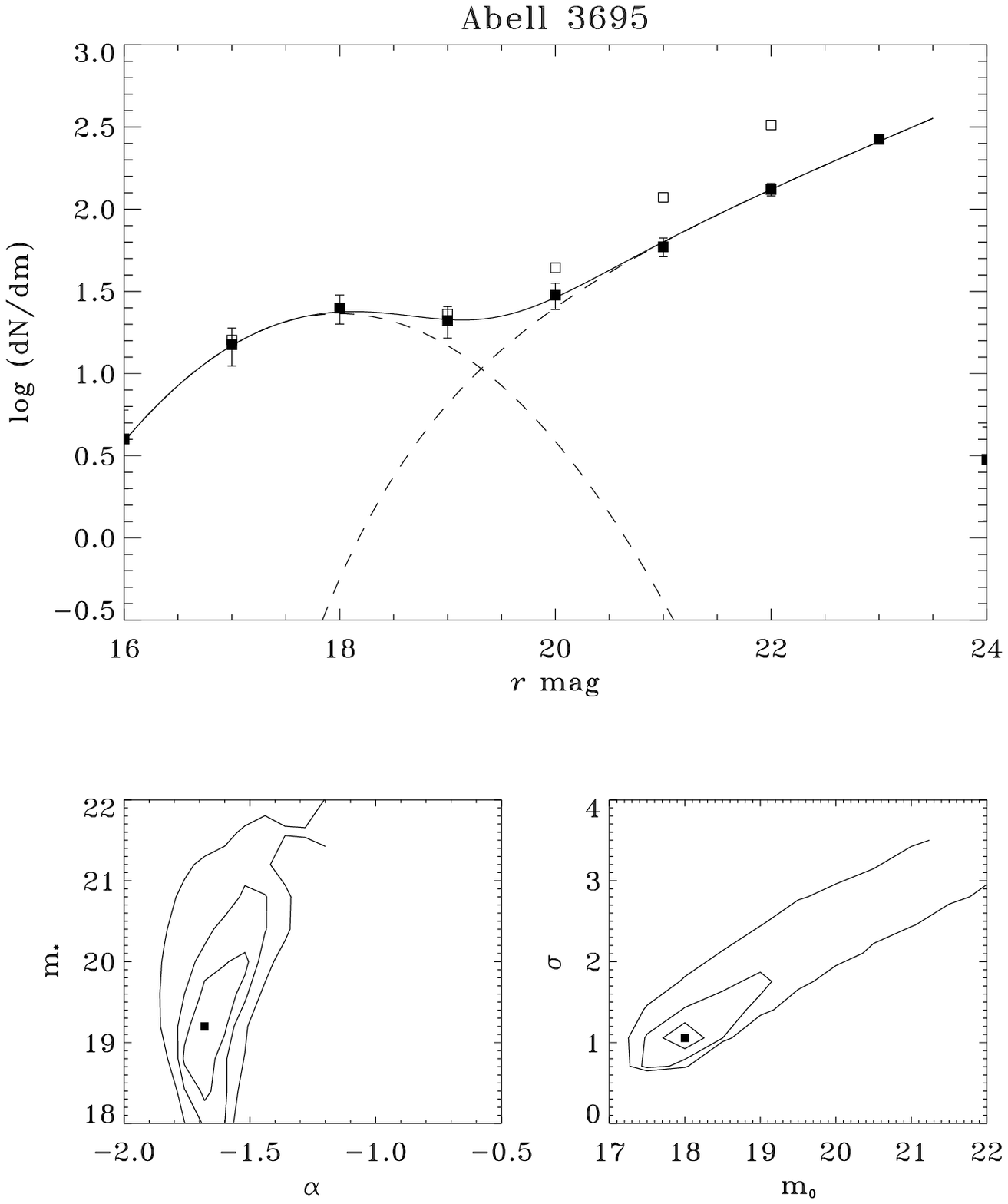} \\
\multicolumn{2}{c}{\includegraphics[width=9cm, height=11.5cm]{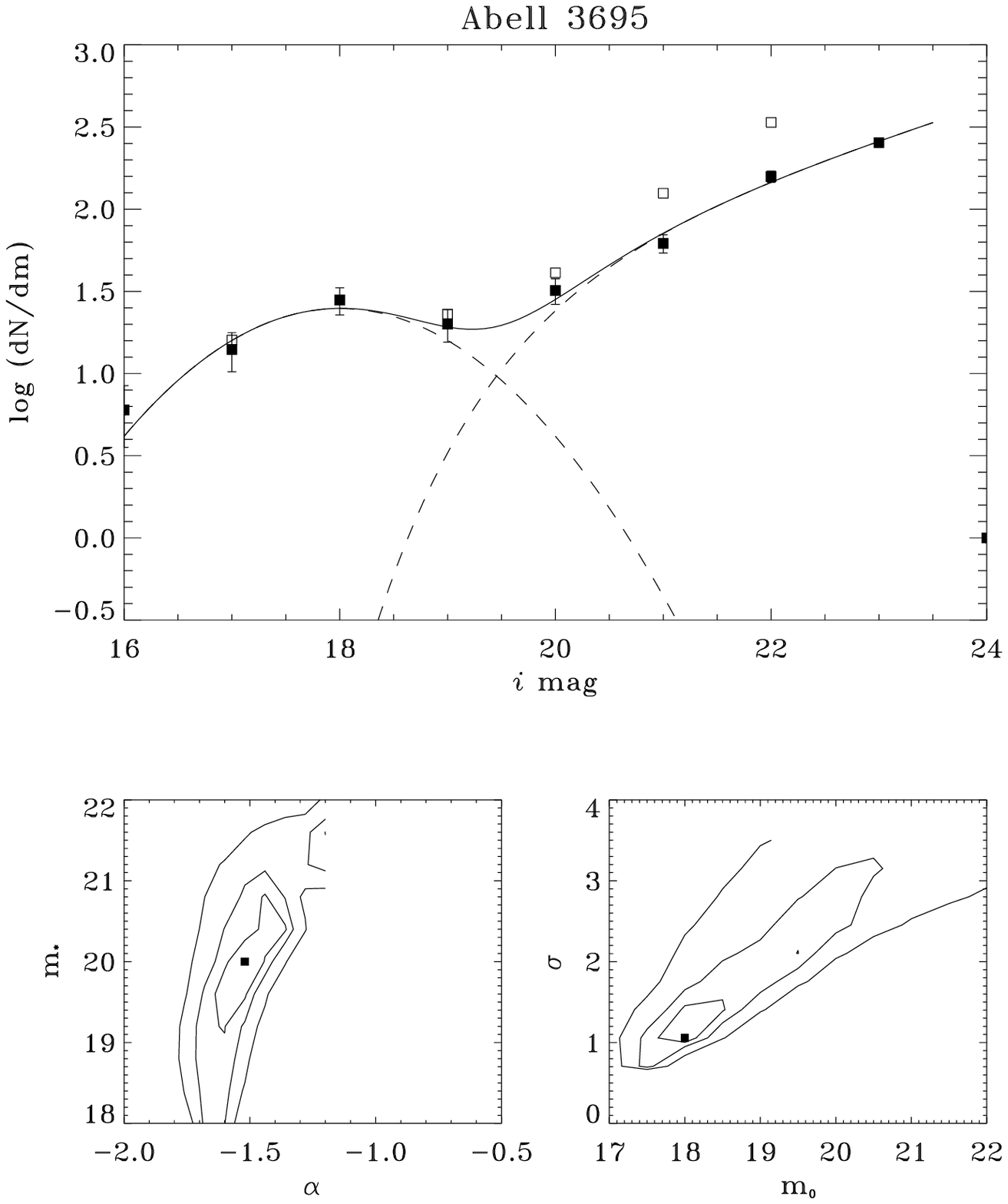}}
\end{tabular}
\caption{A3695}
\label{fig:mxl3695}
\end{figure}

\onecolumn

\begin{figure}[!ht]
\centering
\begin{tabular}{lr}
\includegraphics[width=9cm, height=11.5cm]{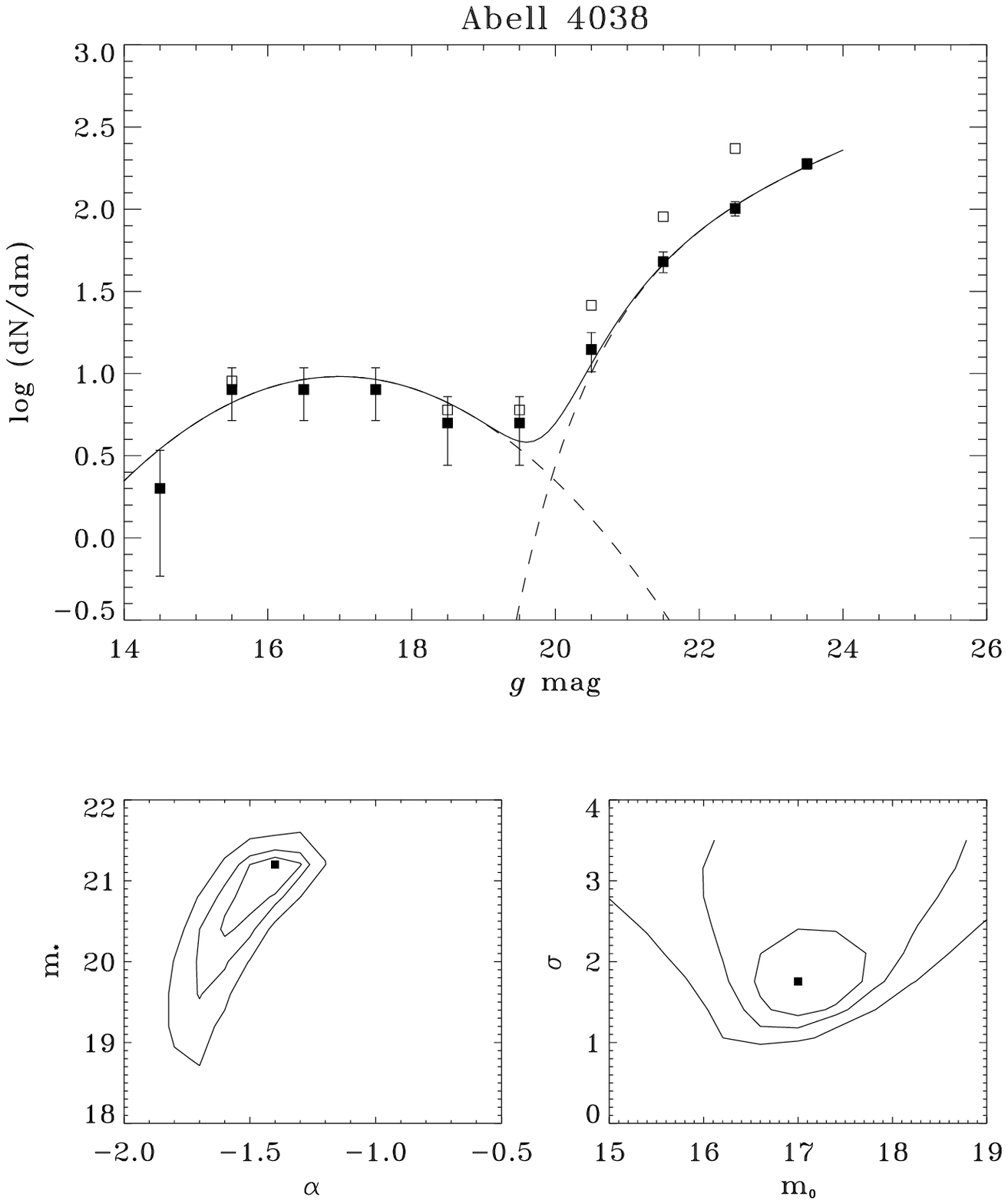} & \includegraphics[width=9cm, height=11.5cm]{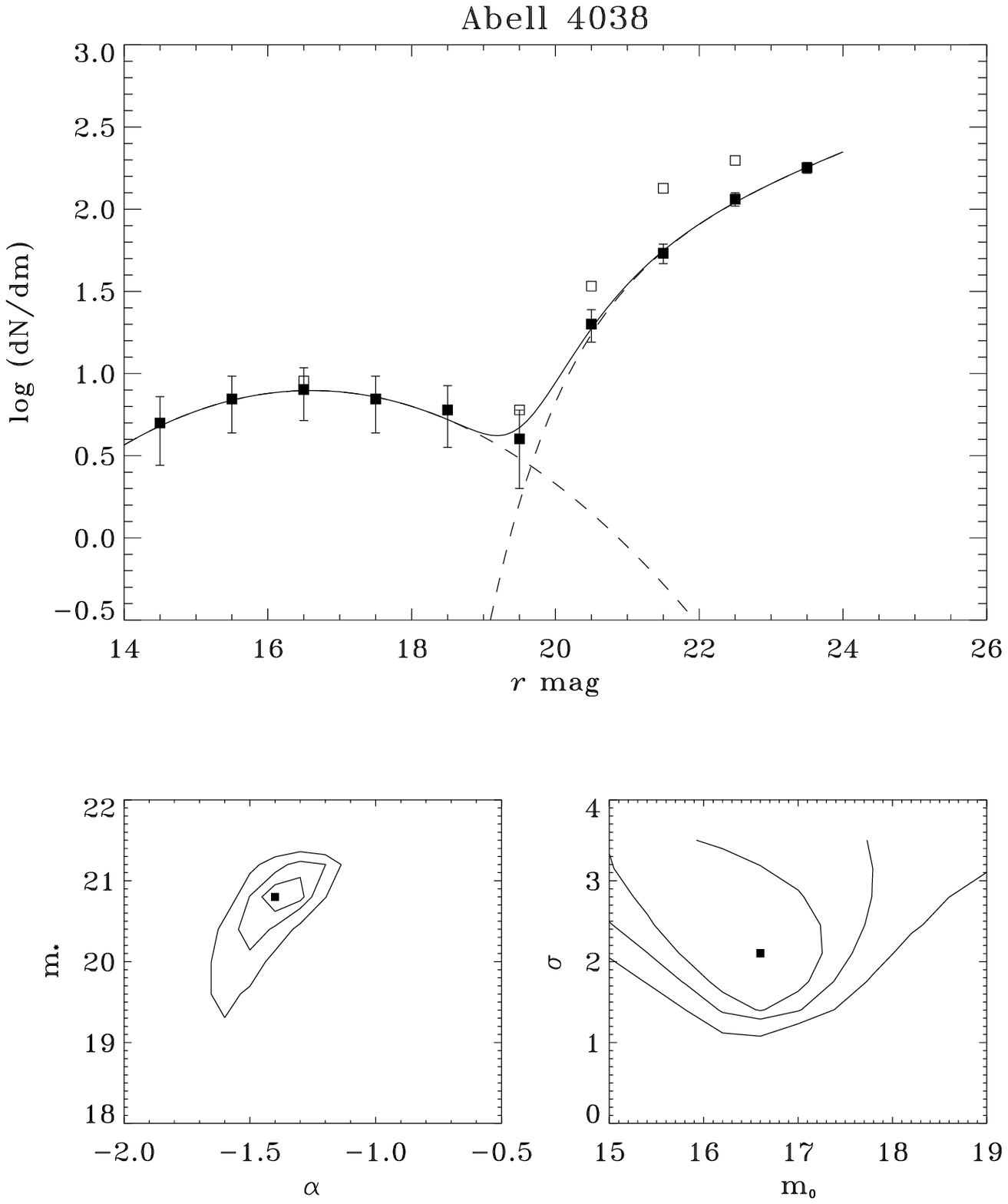} \\
\multicolumn{2}{c}{\includegraphics[width=9cm, height=11.5cm]{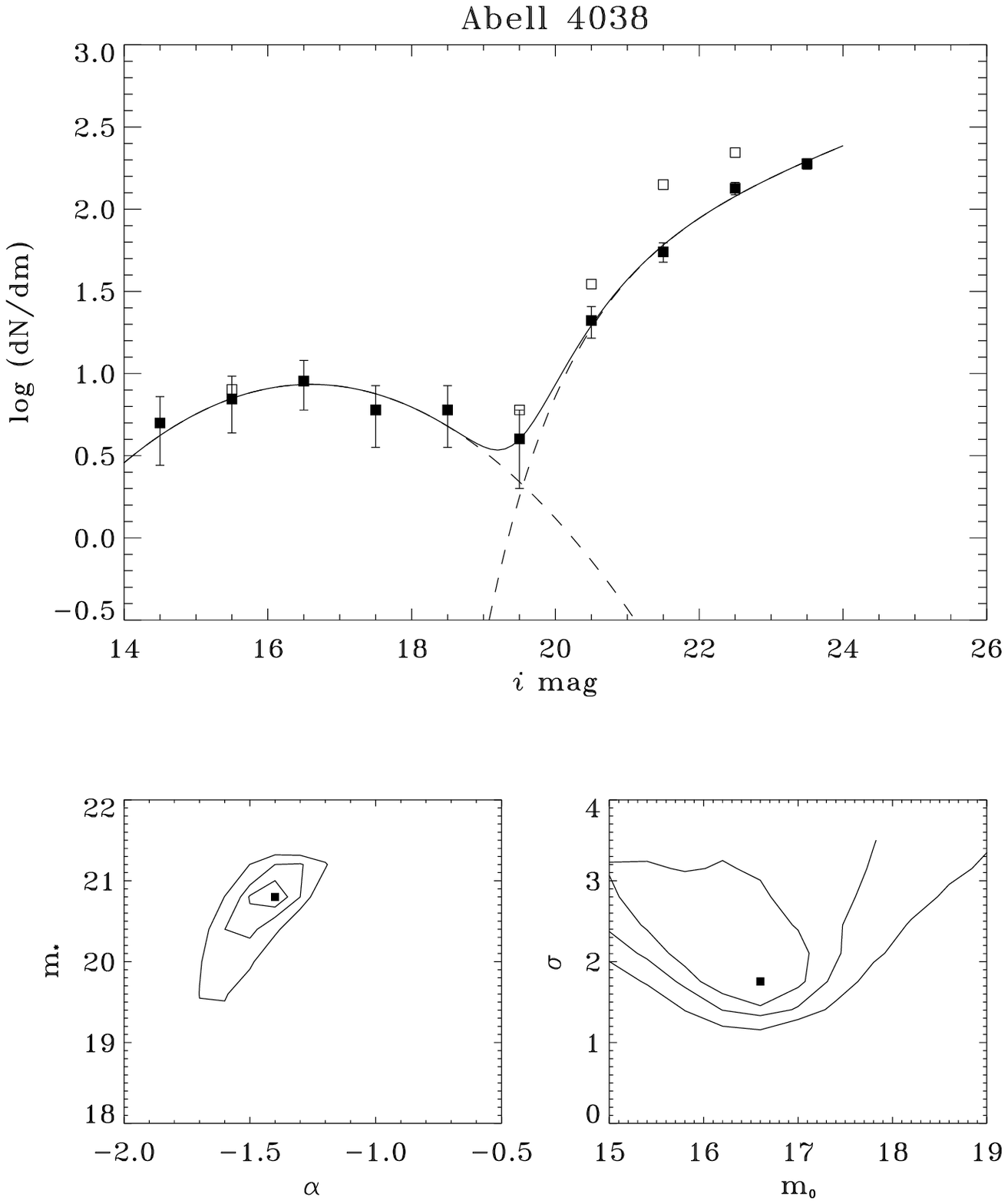}} 
\end{tabular}
\caption{A4038}
\label{fig:mxl4038}
\end{figure}

\onecolumn

\begin{figure}[!ht]
\centering
\begin{tabular}{lr}
\includegraphics[width=9cm, height=11.5cm]{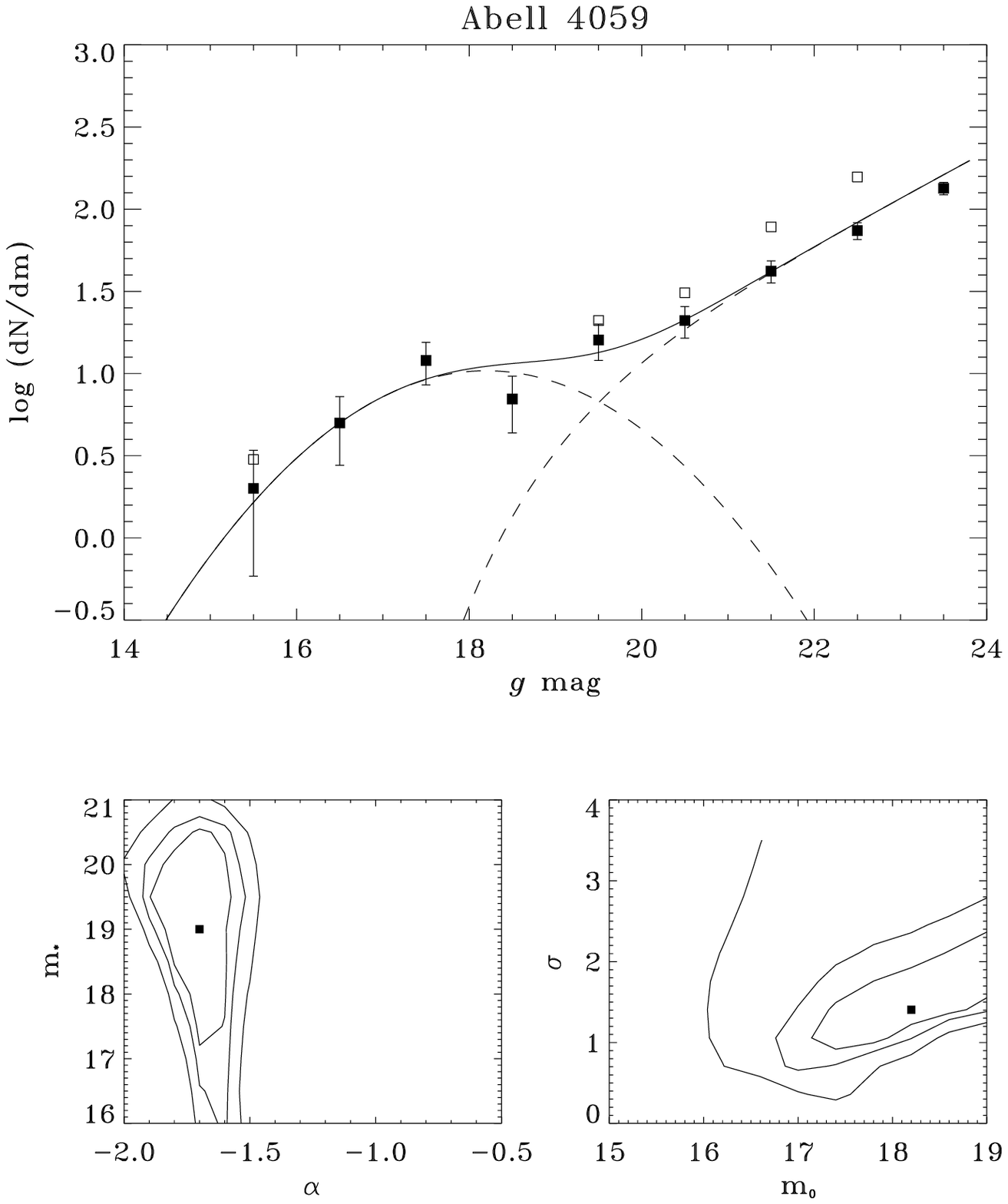} & \includegraphics[width=9cm, height=11.5cm]{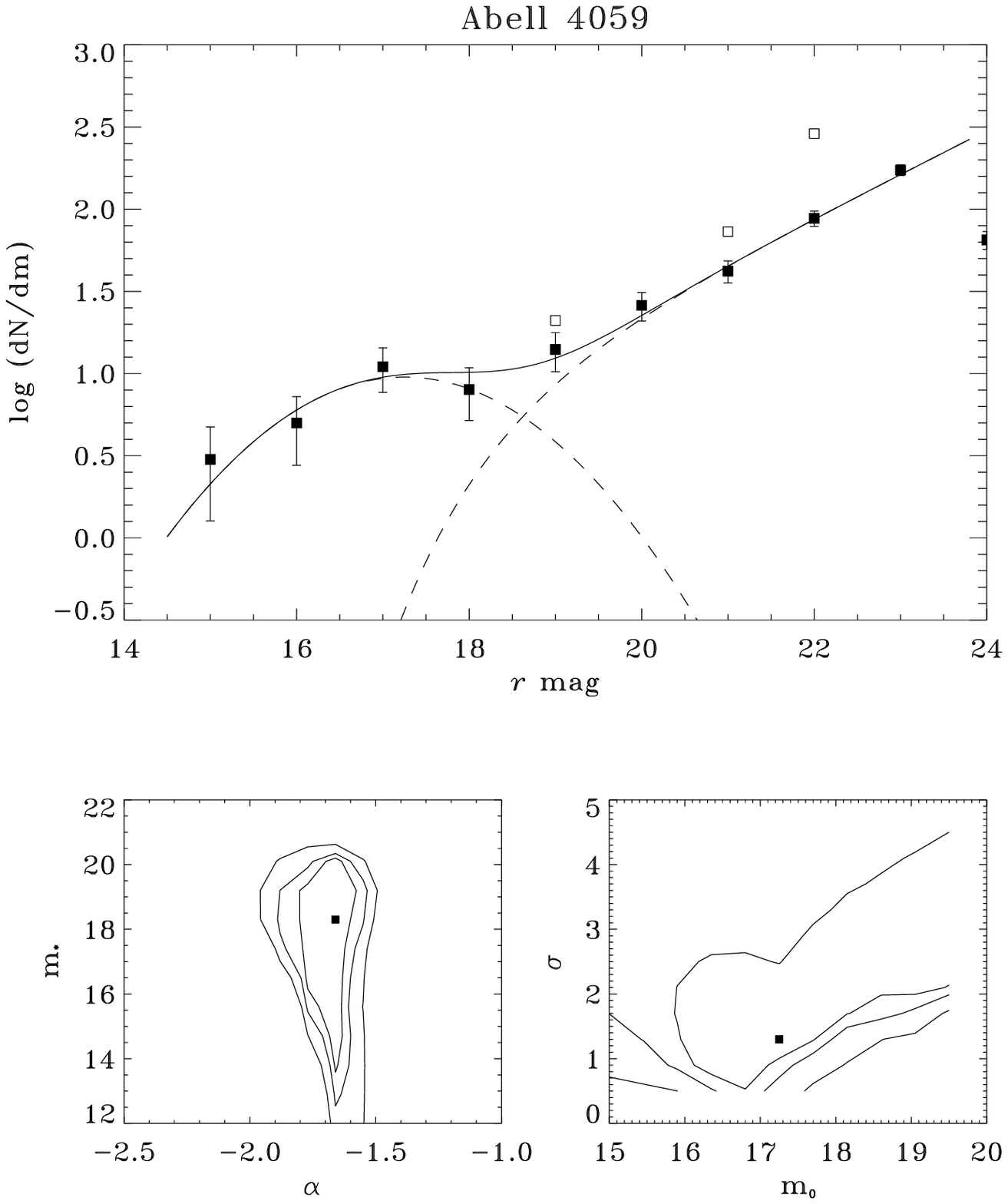} \\
\multicolumn{2}{c}{\includegraphics[width=9cm, height=11.5cm]{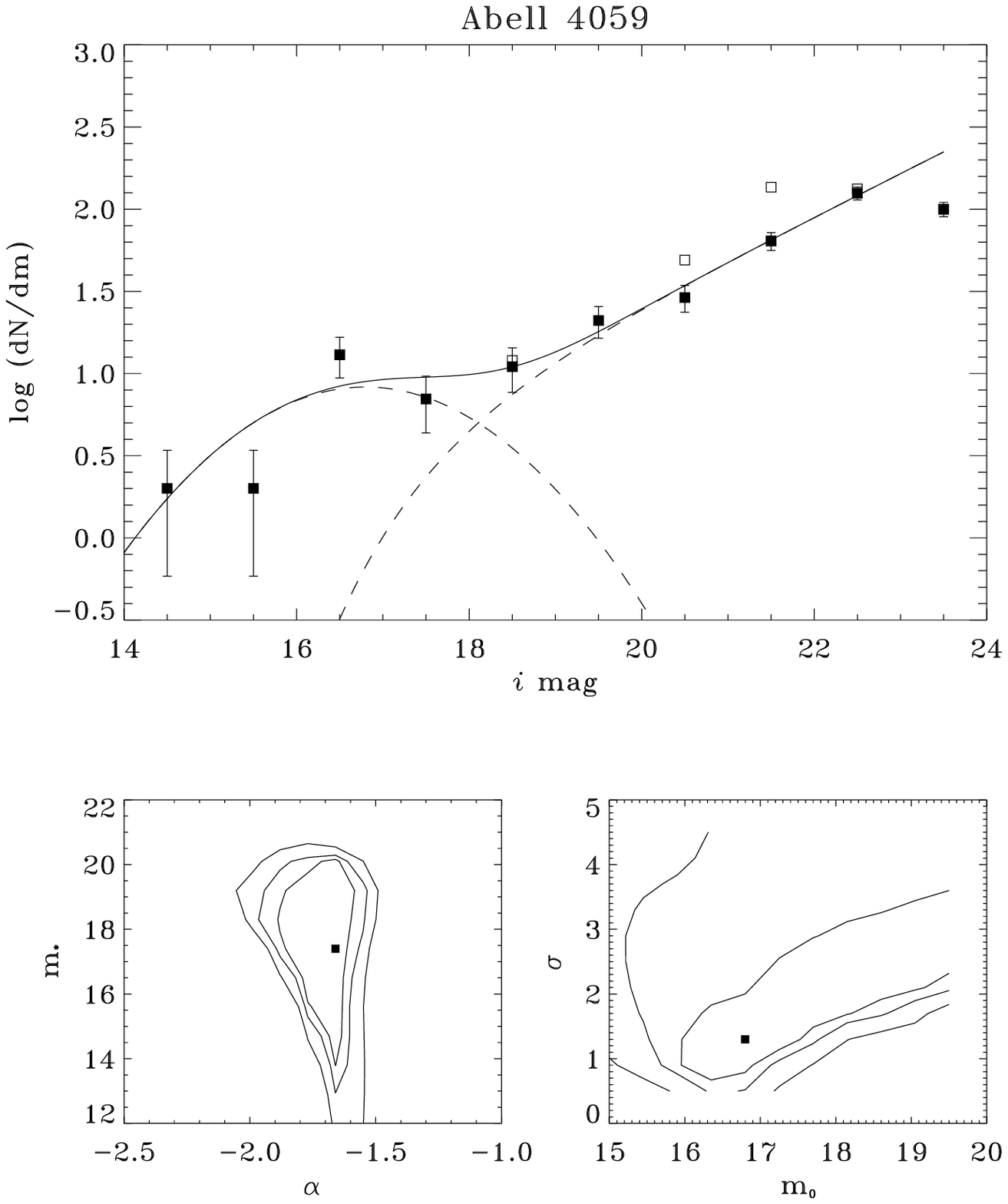}}
\end{tabular}
\caption{A4059}
\label{fig:mxl4059}
\end{figure}

\onecolumn

\begin{figure}[!ht]
\centering
\begin{tabular}{lr}
\includegraphics[width=9cm, height=11.5cm]{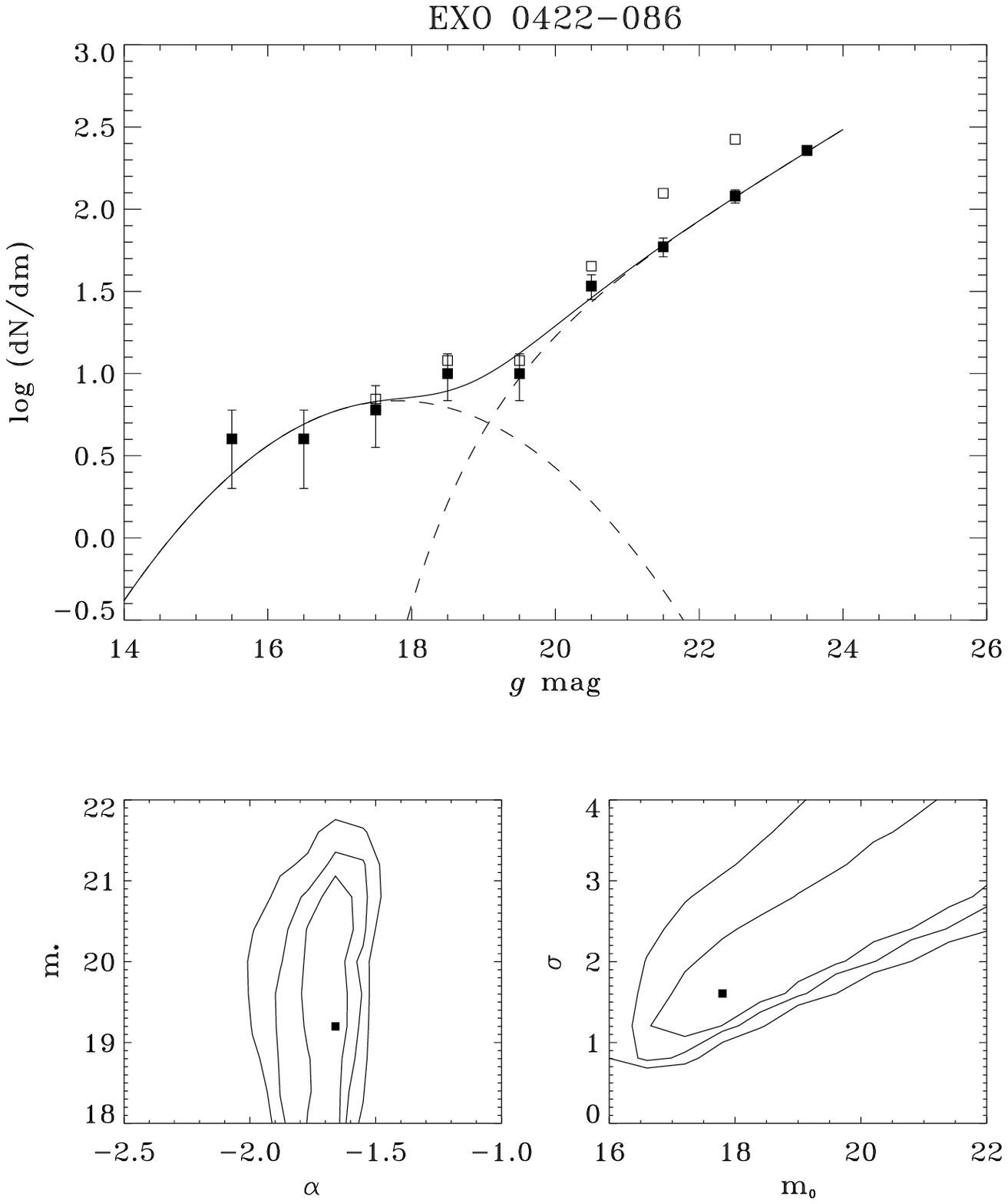} & \includegraphics[width=9cm, height=11.5cm]{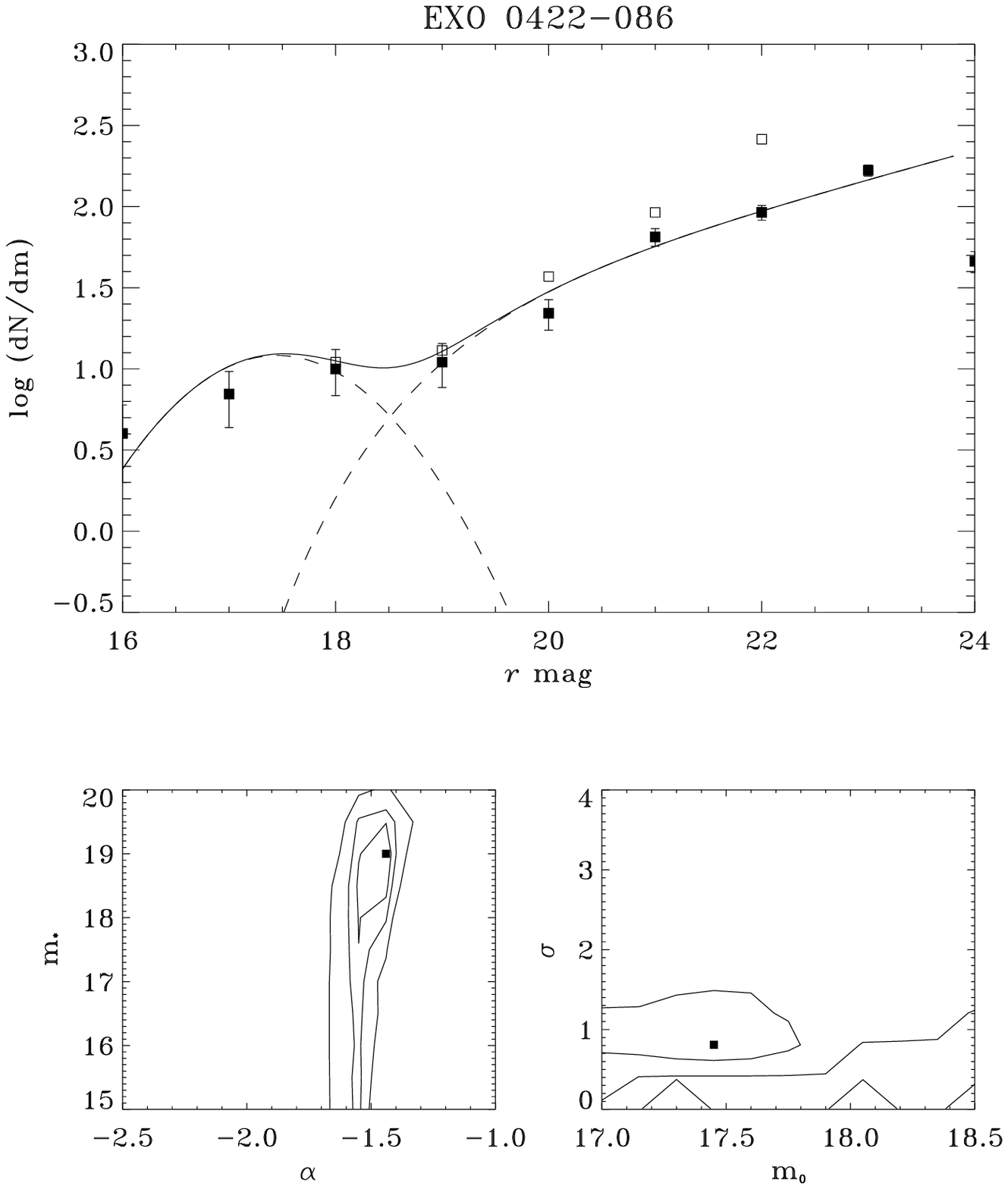} \\
\multicolumn{2}{c}{\includegraphics[width=9cm, height=11.5cm]{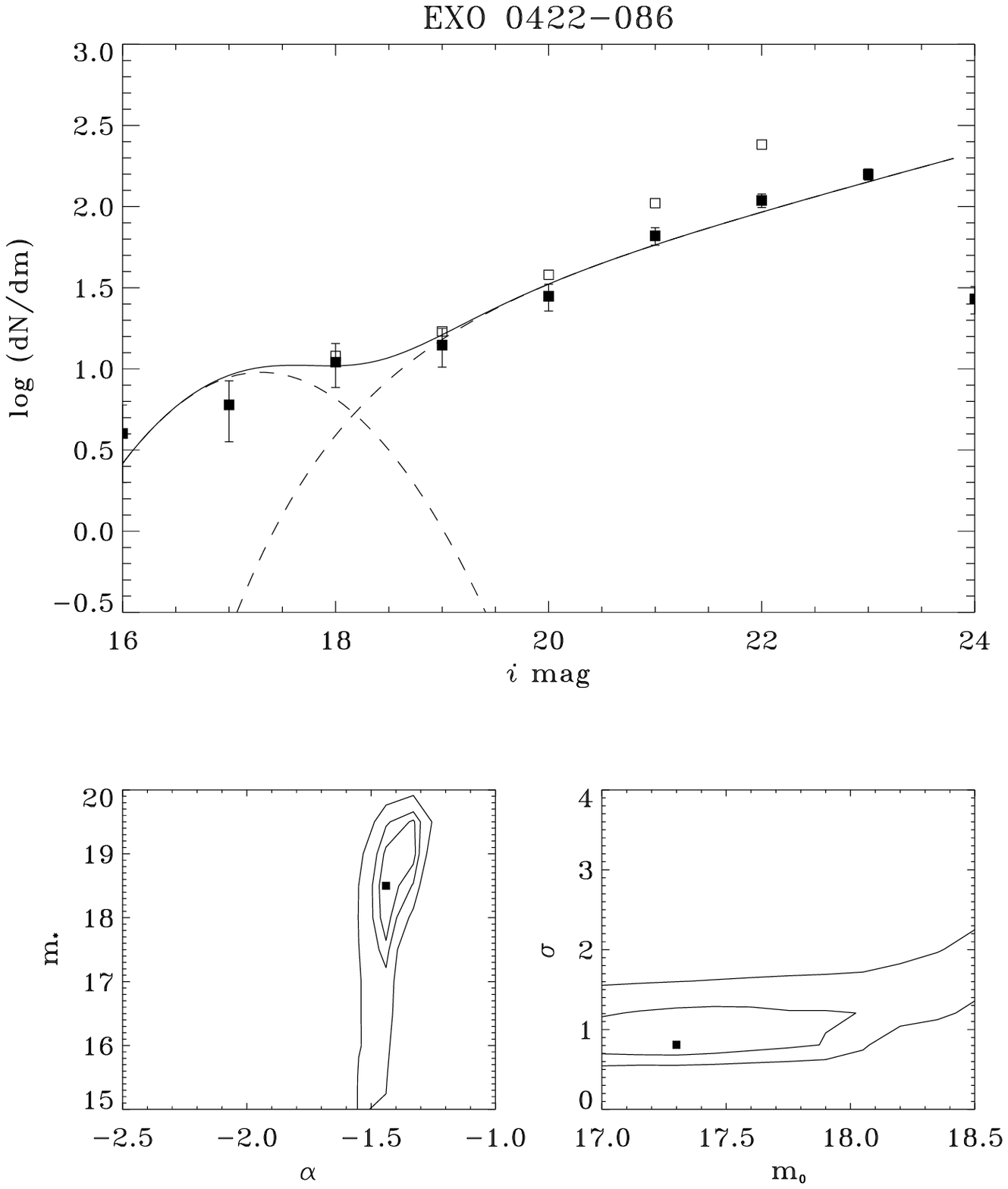}}
\end{tabular}
\caption{EXO 0422-086}
\label{fig:mxlEXO}
\end{figure}

\end{document}